\documentclass[]{MathAppl18}
\usepackage[OT4,T1]{fontenc}
\usepackage[utf8]{inputenc}
\usepackage[usenames,dvipsnames]{xcolor}
\usepackage{polski}
\usepackage{graphicx}
\usepackage{float}
\usepackage{tikz}
\usetikzlibrary{shapes, arrows, positioning}
\usepackage{pgfplots}
\pgfplotsset{width=7cm,compat=1.9}

\usepackage{amsmath}
\usepackage{amssymb}
\usepackage{filecontents}

\usepackage[english,polish]{babel} %PS
\usepackage{lastpage} %PS

\usepackage{etex} %PS

\usepackage{color} 
\usepackage{cite} %PS
\usepackage{mathtools}

\usepackage{subcaption}
\usepackage{placeins}
\usepackage{diagbox}

\usepackage{changes}
\colorlet{cAn}{Bittersweet}
\definechangesauthor[name={Ania}, color=cAn]{Ania}

\colorlet{cBa}{ForestGreen}%Dandelion}
\definechangesauthor[name={Bartek}, color=cBa]{Bartek}

\DeclareMathOperator{\sgn}{sgn}

\RequirePackage[numbers]{natbib}

\usepackage[colorlinks=true]{hyperref}
  \hypersetup{
    pdftitle={Impact of Hill coefficient and time delay on a perceptual decision-making model }, %%<--To wymieni? 
    pdfauthor=Bartek Morawski, %%<--TO wymieni?
    colorlinks,
    urlcolor=blue,
    filecolor=magenta,
    citecolor=green, 
    linkbordercolor={1 1 1}, % set to white
    citebordercolor={1 1 1},  % set to white
    urlbordercolor={ 1 1 1}  % set to white
  } 
 \RequirePackage[hyperpageref]{backref} 
    \renewcommand*{\backref}[1]{}  
    \renewcommand*{\backrefalt}[4]{
       \ifcase #1 
          No cited.
       \or
          Cited on p. #2.
       \else
          Cited on pp. #2.
       \fi}  
\newcommand{\orcid}[1]{\href{https://orcid.org/#1}{\includegraphics[scale=.05]{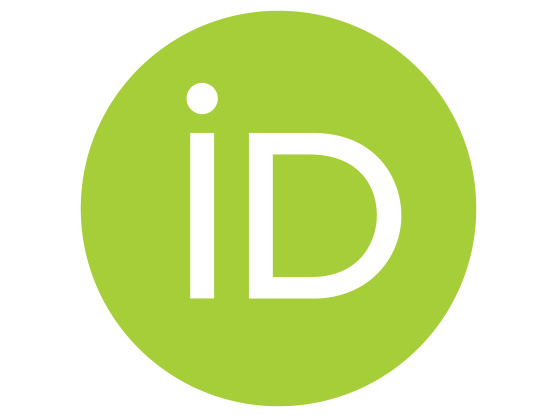}}}

\def\repo{http://wydawnictwa.ptm.org.pl/index.php/matematyka-stosowana/article/viewArticle}  
 
\pdfoutput=1
\usepackage{graphicx}
\usepackage{wrapfig}
\graphicspath{{./Figures/},{./Pictures/}}
\usepackage{multicol}
\firstpage{i}    
\LogoG{\includegraphics[width=0.18\textwidth]{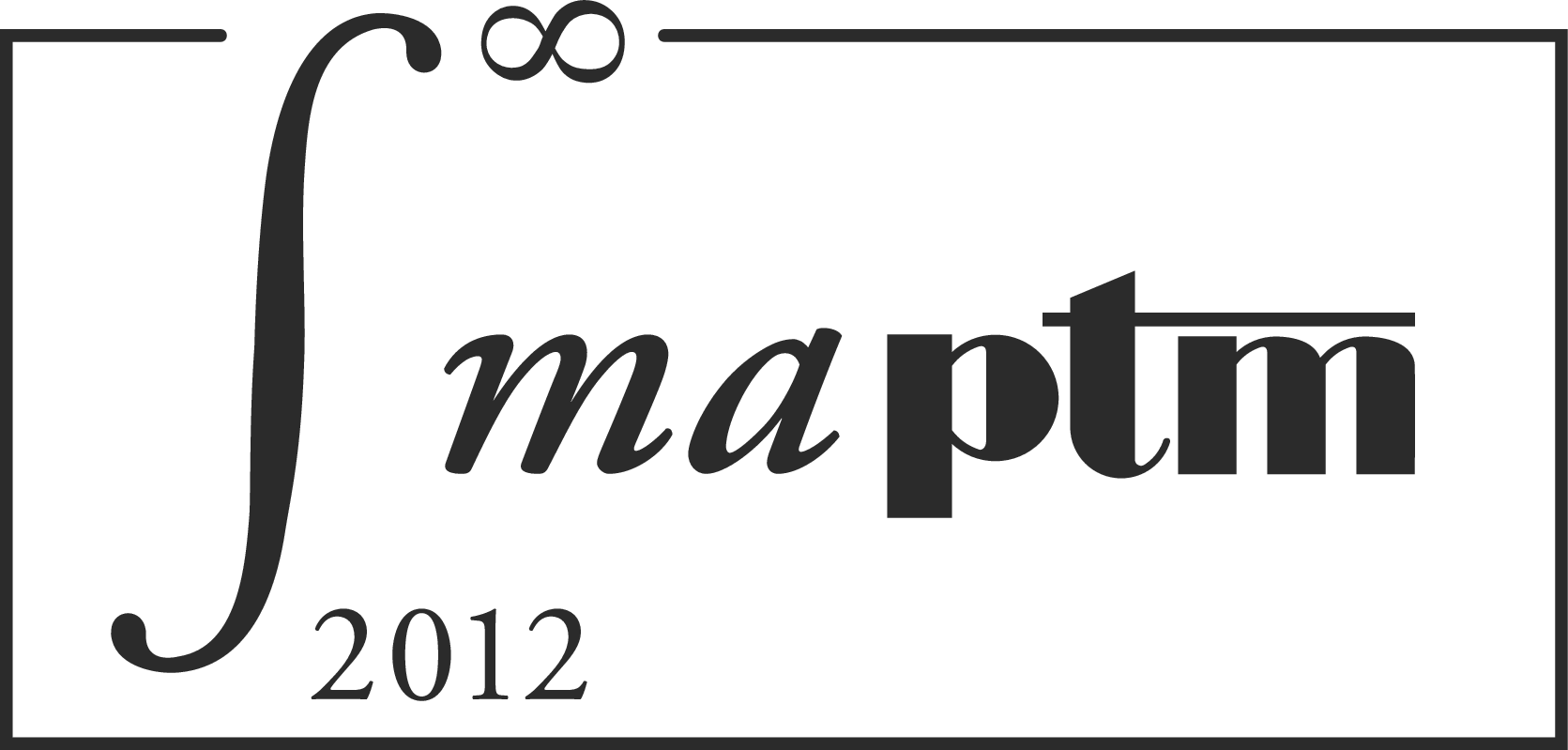}}
\volume{46}
\fasc{2}
\years{2018}
\LogoGMS{\includegraphics[width=0.18\textwidth]{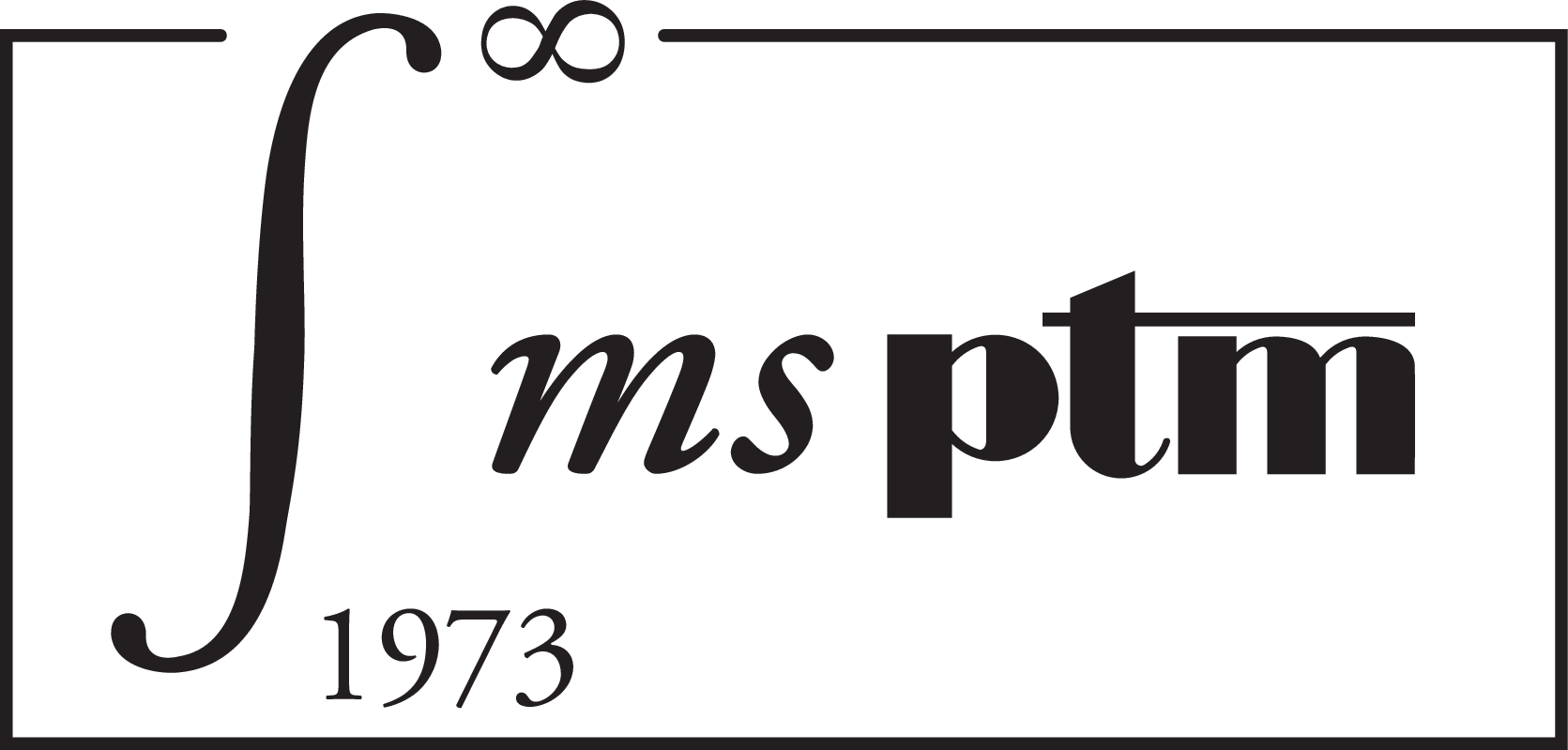}}
\volumeMS{26}
\numberMS{63}
\wwwtrue
\secnameMS{MATHEMATICS FOR SOCIAL AND ART SCIENCES}
\pages{193--206}
\received[5th of June 2017]{ 11th of January 2017}
\lastrevision{}
\logo{}{}{}%Do not change 9 above lines
\def\doinum{10.14708/ma.v43i1.xxx}%% http://wydawnictwa.ptm.org.pl/index.php/matematyka-stosowana/article/view/614
\title[Mathematica Applicanda Template]{Impact of Hill coefficient and time delay on a perceptual decision-making model 
}
\tpauthortrue % PS11lis12
\author[B. Morawski]{Bartłomiej Morawski \orcid{0009-0003-2839-8575}}%(Albuquerque)
\affiliation{University of Warsaw}
\email{bm417744@students.mimuw.edu.pl}
\author[A. Czartoszewska]{Anna Czartoszewska \orcid{0009-0005-9524-284X}}%(Albuquerque)
\affiliation{University of Warsaw}
\email{a.czartoszewsk@student.uw.edu.pl}

\begin{document}
\vspace{-5ex}
%\Poczatek
%\Chapter
%\pagenumbering{roman}
% \setcounter{page}{193} %%This command starts the numerations of pages
\selectlanguage{english}\Polskifalse
%\selectlanguage{polish}\Polskitrue

\begin{abstract}
In this paper, a neural mass perceptual decision making model introduced by Piskała et al. \cite{piskala2017neural} is analyzed. The model describes activity of two neuron populations influenced by each other and external inputs. The groups' activities correspond to the process of making a perceptual binary decision. Existing results are generalized by investigating the impact of both a delay in self-inhibition and a generic Hill coefficient on solutions to the system of differential equations. Several versions of the model with various assumptions are compared using analytical and numerical methods.
\end{abstract}
%\tableofcontents

\section{Introduction}

\subsection{Neuroscientific context}

\noindent Decision making is one of the most important functions of the brain, enabling humans and other animals to function in everyday life. Apart from complex, conscious decisions, our brain makes many perceptual decisions that we may not be aware of, concerning observed shapes, colors, direction of movement, location of a sound source, and more. The decision is based on the activity of the corresponding neural circuits. The difficulty of the task arises from the fact that the neural signals elicited by the stimuli are disrupted by noise coming from various sources \cite{smith2009ideal}. In this paper, we consider noise coming from the `background' activity of the brain, i.e. the activity not related to the task, and the mutual stimulation through synaptic connections between neural populations. 

In laboratory conditions, perceptual decision making is commonly studied using tasks in which a participant must choose between two available responses when perceiving two kinds of stimuli differing in one parameter (two-alternative forced choice, \cite{smith2009ideal}). An example of such an experiment is the one described in an article by Shadlen \& Newsome \cite{shadlen1996motion}, conducted on monkeys. The task was to recognize the direction of movement of the majority of dots displayed on a screen. This type of research has allowed for the identification of the neural basis of decision-making processes.

Given the importance of decision making and the goal of understanding human functioning, developing mathematical models of these processes can provide valuable insights. They can help us understand how a given process occurs in reality, making it possible to confront theories with real data. It can also give us ideas about the problems that may arise within the process. In particular, a well-constructed model can help identify potential sources of dysfunction and guide the development of therapeutic interventions.

Based on Shadlen \& Newsome \cite{shadlen1996motion} and similar studies, Xiao-Jing Wang \cite{wang2002probabilistic} developed a mathematical model that describes the dynamics of activity of two groups of neurons, with synaptic connections, that respond to two different types of stimuli (movement to the right and movement to the left). The system's task is to correctly identify the type of stimulus by assessing which of the groups was initially stimulated. This model is based on the winner-take-all mechanism, which means that the decision made must be binary and is made in favor of the population with greater activity. In the case of no difference, no decision can be made. This assumption can be implemented using various decision criteria, which we discuss later.

The model analyzed in this paper also takes into account short-term synaptic plasticity (STP, \cite{ermentrout2010mathematical}). There are two types of STP: synaptic depression and synaptic facilitation. Both of these states signify a change in the `ease' of excitement of a cell. Depending on the properties of the synapse and the neuron itself, successive excitatory signals have a decreasing (in the case of depression) or increasing (in the case of facilitation) influence on the cell, altering its sensitivity to incoming impulses. We take into account only the facilitation.

\subsection{Presentation of the model}

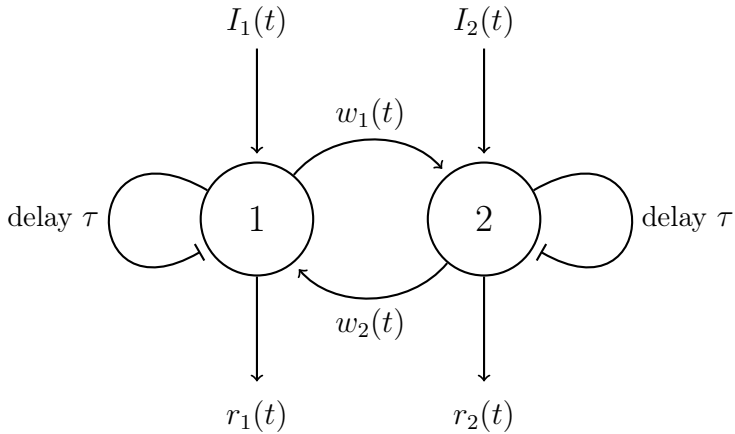
\begin{figure}[ht]
\centering
    \begin{tikzpicture}[node distance=1.5cm and 1.5cm, line width = 0.8]
    \node[circle, draw, minimum size = 1.5cm](a){\Large 1};
    \node[circle, draw, minimum size = 1.5cm][right=of a](b){\Large 2};
    \node[below=of a] (r1) {\large $r_1(t)$};
    \path[->] (a) edge[shorten >=3pt] (r1);
    \node[below=of b] (r2) {\large $r_2(t)$};
    \path[->] (b) edge[shorten >=3pt] (r2);
    \node[above=of a] (Is) {\large $I_1(t)$};
    \path[->] (Is) edge[shorten >=3pt] (a);
    \node[above=of b] (I) {\large $I_2(t)$};
    \path[->] (I) edge[shorten >=3pt] (b);
    \path (a) edge [in=210,out=150,distance=2cm, -Bar, shorten >=3pt] node[midway, left]{delay \large $\tau$}(a);
    \path (b) edge [in=330,out=30,distance=2cm, -Bar, shorten >=3pt] node[midway, right]{delay \large $\tau$} (b);
    \path[->] (a) edge [bend left=50, shorten >=3pt] node[above]{\large $w_1(t)$} (b);
    \path[->] (b) edge [bend left=50, shorten >=3pt] node[below]{\large $w_2(t)$} (a);
    
    \end{tikzpicture}
\caption{Scheme of the modeled system.}
\label{fig: schemat}
\end{figure}

\noindent The winner-take-all mechanism mentioned earlier also underlies the decision making model proposed by Piskała et al. \cite{piskala2017neural} and further extended and analyzed by Bielczyk et al. \cite{bielczyk2019time} and Foryś et al. \cite{forys2017impact}, which is the subject of analysis in this work. The scheme of the system is depicted in Figure \ref{fig: schemat}. The model describes the dynamics of the network, that is, signals generated by populations, denoted as $r_1(t)$ and $r_2(t)$, which correspond to the two decision options. The two populations receive certain inputs, marked by $I_1(t)$, $I_2(t)$, which include the signal coming from the stimulus as well as the background noise. Additionally, the populations are influenced by each other through plastic synapses. The decrease of the activation is ensured by self-inhibition, which can be time-delayed by $\tau$. Introducing the delay was shown to explain some impairments in the decision-making process (making wrong decision and ambiguity; \cite{bielczyk2019time}, \cite{forys2017impact}).

In the analyzed model, synaptic plasticity of the connections between the populations is incorporated using differential equations describing changes in  connection weights, denoted by $w_1(t)$ and $w_2(t)$. Their magnitudes are influenced by the activity of the populations, and the maximum strength of a synaptic connection is anatomically limited, with the maximum denoted by $\epsilon$. We assume that the weight of a connection decreases exponentially in the absence of activity and that its increase (synaptic facilitation) depends on the product of the signals of both neuronal masses through the Hill sigmoid function: $f_n(r_1r_2) = \frac{(r_1r_2)^n}{1+(r_1r_2)^n}.$ The Hill coefficient $n$ is usually a positive natural number. In the original article \cite{forys2017impact}, the authors adopted $n = 2$ due to the biological plausibility of this value. In this work, we investigate how varying values of the parameter affect the model's behavior. We therefore consider $n \in [1, +\infty)$, since there are no significant qualitative changes in the behavior of the function $f_n$ when $n>1$. The only exception is $n=1$, as seen in Figure \ref{fig: hill}.

\pgfplotsset{width=6cm,compat=1.7}
\begin{figure}[H]
\centering
\begin{tikzpicture}
\begin{axis}[
    xmin = 0,
    ymin = 0,
    xmax = 4,
    ymax = 1,
    xlabel = \(r_1r_2\),
    ylabel = {\(f_n(r_1r_2)\)},
    legend pos=south east,
]
%n=1
\addplot [
    domain=0:5, 
    samples=100, 
    color=red,
]
{x/(1+x)};
\addlegendentry{\(n = 1\)}
%n=2
\addplot [
    domain=0:5, 
    samples=100, 
    color=orange,
]
{x^2/(1+x^2)};
\addlegendentry{\(n = 2\)}
%n=3
% \addplot [
%     domain=0:5, 
%     samples=100, 
%     color=yellow,
% ]
% {x^3/(1+x^3)};
% \addlegendentry{\(n = 3\)}
%n=4
\addplot [
    domain=0:5, 
    samples=100, 
    color=green,
]
{x^4/(1+x^4)};
\addlegendentry{\(n = 4\)}
%n=10
\addplot [
    domain=0:5, 
    samples=100, 
    color=blue,
]
{x^10/(1+x^10)};
\addlegendentry{\(n = 10\)}

\end{axis}
\end{tikzpicture}
\caption{Shape of the Hill function $f_n$ depending on the parameter $n$.}
\label{fig: hill}
\end{figure}
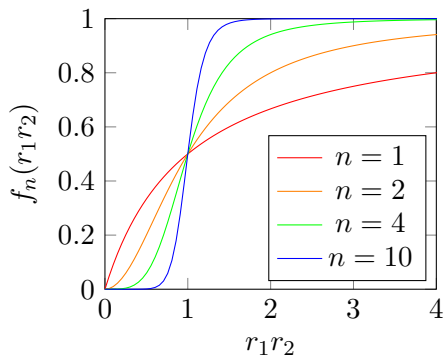

The system of equations corresponding to the described situation is
\begin{equation}\label{eq:all4}
    \begin{aligned}
        \tau_r \dot{r}_1 &= -r_1(t-\tau) + w_2r_2+I_1(t) , \\
        \tau_r \dot{r}_2 &= -r_2(t-\tau) + w_1r_1+I_2(t) , \\
        \tau_w \dot{w}_1 &= -w_1 + \epsilon f_n(r_1r_2) , \\
        \tau_w \dot{w}_2 &= -w_2 + \epsilon f_n(r_1r_2) .
    \end{aligned}
\end{equation}

\noindent While in general the inputs $I_1(t), I_2(t)$ are time-dependent, we assume them to be constant in time, making the equations autonomous. The coefficients $\tau_r$ and $\tau_w$ are the time scales of activity and plasticity, respectively. All variables, coefficients, and inputs $I_{1, 2}$ are positive, and we are interested in solutions with positive variables $r_{1,2}, w_{1,2}$ (for biological interpretability). For positive initial conditions in the non-delayed case, the variables remain positive as well. In the non-zero delay case positivity of solutions is not guaranteed, which we show in numerical simulations. In such situations, the biological sense of the model is lost.

% In this study, we analyse a few variants of the original model, which are based on various assumptions. We consider the interaction between the values of the parameters (especially $n$ in $f_n$ and delay $\tau$) and the behavior of the models. In particular we show changes in the number and stability of steady states. We conclude that $n$ and $\tau$ can change the qualitative behavior of the system. Moreover, we conduct numerical simulations showing the impact of non-symmetric inputs (stimulus present) and the changes introduced by varying parameters.

\par
\noindent In this study, we analyze a few variants of the original model, each based on different assumptions. We investigate how the interaction between model parameters -- particularly the Hill coefficient $n$ in $f_n$ and the delay $\tau$ -- affects system behavior. In particular, we demonstrate how variations in these parameters influence the number and stability of steady states, showing that both $n$ and $\tau$ can induce qualitative changes in the system's dynamics. Additionally, we present numerical simulations that illustrate the effects of asymmetric inputs (an applied stimulus) and further explore how varying parameters modify the system's behavior.

% \subsection{Quasi-steady state approximation}

% \noindent \aniaa{In this study, we will investigate how this parameter affects the existence and stability of stationary states and the perceptual decision made by the system. (??? Tak było u mnie, ale tutaj to nie wiem)}

% \bartek{w tym study zbadamy sobie jak interplay miedzy zmianami parametrow i roznymi warianty modelu wychodzace z roznych zalozen przekładają sie na rozwiazania, co jest podobne a co jak sie zmienia i pod wplywem czego. no i zobaczymy że pewne wlasnosci (istnienie i liczba stanow stacjonarnych) przy pewnych zmianach pozostaja stale, podczas gdy stabilnosc moze sie zmieniac no i że n i $\tau$ zmieniaja jakosciowo zachowania}

% \hline

% \aniaa{ To wywaliłam:
% The first two equations describe the firing rates $r_i$ of two groups, referred to as group $1$ and $2$ respectively. External inputs \bartekk{background activity?} are marked by $I_1(t), I_2(t) $ and while in general they are time-dependent, in further symbolic analysis we assume them to be constant in time, making the equations autonomous. The weights $w_i$ are connected to the latter two equations, which are accountible for short-term plasticity \bartekk{TODO cytowanie} -- the strengths of groups' influences on each other are functions of time.}

\section{Model analysis}
% \bartek{do zmiany ponizej}

\noindent In this chapter, we investigate two simplified versions of System \eqref{eq:all4}. The first one is a quasi-steady state approximation, which reduces the original four equations to two equations for firing rates $r_{1,2}$.  The other one allows the weights to not be in their steady state, but assumes them to be equal, resulting in a three-equation system. 
\par
Note that the systems of DDEs in this section might produce solutions which are negative at some point, contrary to zero-delay System \eqref{eq:all4bezd}. In such cases the models lose their biological sense, which is a limitation to be considered. However, we are mostly concerned with solutions' behavior near positive steady states.

\subsection{Model with quasi-steady state approximation and zero delay}
% \aniaa{Zastanawiam się czy tu powinno być 'the' w tytule, tak samo w 2.3.}

\noindent We begin by focusing on the model with zero-delay self-inhibition, for which the equations are
\begin{equation}\label{eq:all4bezd}
    \begin{aligned}
        \tau_r \dot{r}_1 &= -r_1 + w_2r_2+I_1, \\
        \tau_r \dot{r}_2 &= -r_2 + w_1r_1+I_2, \\
        \tau_w \dot{w}_1 &= -w_1 + \epsilon f(r_1r_2), \\
        \tau_w \dot{w}_2 &= -w_2 + \epsilon f(r_1r_2).
    \end{aligned}
\end{equation}

\noindent As the plasticity is short-term, we assume that the time scale $\tau_w$ for weights is much smaller than the one for firing rates, $\tau_r$. This means that the weights change faster than the rates, and therefore can be assumed to be in their steady states at all time, significantly simplifying the analysis. Note that the quasi-steady state assumption does not mean that the weights are constant in time, since they depend on the rates $r_{1,2}$ through the Hill function.
\par
The solutions to System \eqref{eq:all4bezd} exist on interval $[0, T]$ for some $T>0$ and are unique, because its right-hand side is of class $C^1$. The solutions with a positive initial condition remain positive. We perform quasi-steady state approximation, assuming the weights $w_{1,2}$ to be at equilibrium: $w_{1,2} = \epsilon f(r_1r_2)$.
% $$
% \begin{aligned}
%     \dot{w_i} &= 0, \\
%     \iff\ \, 0 &= -w_i + \epsilon f(r_1r_2), \\
%     \iff w_i &= \epsilon f(r_1r_2) .
% \end{aligned}
% $$
% \textcolor{red}{TODO poprawki} 
Moreover, from the form of the right-hand sides of the System \eqref{eq:all4bezd} we can conclude that all variables are positive and $w_1, w_2 < \epsilon$.
%The derivatives for the weights $\dot{w}_i$ have an upper bound $\dot{w}_i \leq -w_i + \epsilon$. since the function $f$ takes the values in the interval $(0, 1)$ for positive arguments. The solutions to the equations $\dot{w} = -w + \epsilon$ monotonically tend to $\epsilon$, so $w \leq \max \{w(0), \epsilon\}$. This in turn means that $w_i \leq \max \{ w_i(0), \epsilon \}$ for all $t>0$ and $i \in \{ 1, 2\}$. Denoting $w_{\max} = \max \{ w_1(0), w_2(0), \epsilon \}$ we can estimate
The solutions can be extended for all $t > 0$, because $w_1$ and $w_2$ do not converge to 0 or $\epsilon$ for any $t$ and the growth of the rates $r_1, r_2$ is at most exponential, which can be concluded from the following estimations:

\begin{equation*}
\begin{aligned}
    \dot{r}_1 + \dot{r}_2 &= \frac{1}{\tau_r}(-r_1 + w_2r_2+I_1 -r_2 + w_1r_1+I_2 ) \\ 
    &\leq \frac{1}{\tau_r} \big( (\epsilon-1) (r_1+r_2) + I_1 + I_2 \big) .
\end{aligned}
\end{equation*}
%which means the growth of the rates $r_i$ is at most exponential. Therefore, the solutions exist for all $t>0$. \bartek{część wyżej myślę że będzie do skrócenia} 
% \ania{Przerobiłam to co było wyżej, żeby było krócej (wzorując się na moim streszczeniu konferencyjnym). Zostawiłam Twoje rzeczy w komentarzach.}
\par
\begin{figure}[h]
    \centering
    \begin{subfigure}[b]{0.45\linewidth}
        \centering
        \includegraphics[width=\linewidth]{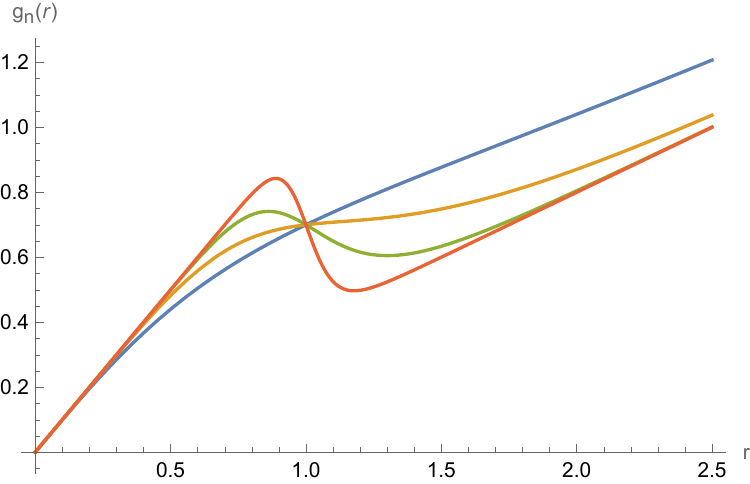}
        \caption{}
        \label{fig:gn_e06}
    \end{subfigure}
    \hfill
    \begin{subfigure}[b]{0.45\linewidth}
        \centering
        \includegraphics[width=\linewidth]{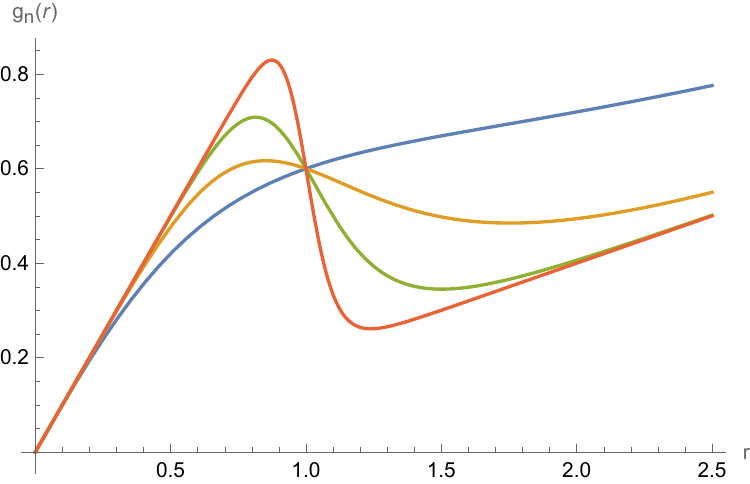}
        \caption{}
        \label{fig:gn_e08}
    \end{subfigure}
    \vfill
    \begin{subfigure}[b]{0.45\linewidth}
        \centering
        \includegraphics[width=\linewidth]{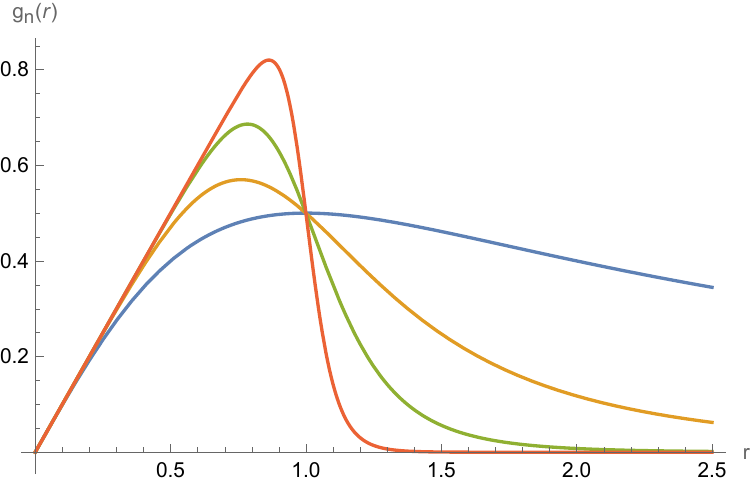}
        \caption{}
        \label{fig:gn_e1}
    \end{subfigure}
    \hfill
    \begin{subfigure}[b]{0.45\linewidth}
        \centering
        \includegraphics[width=\linewidth]{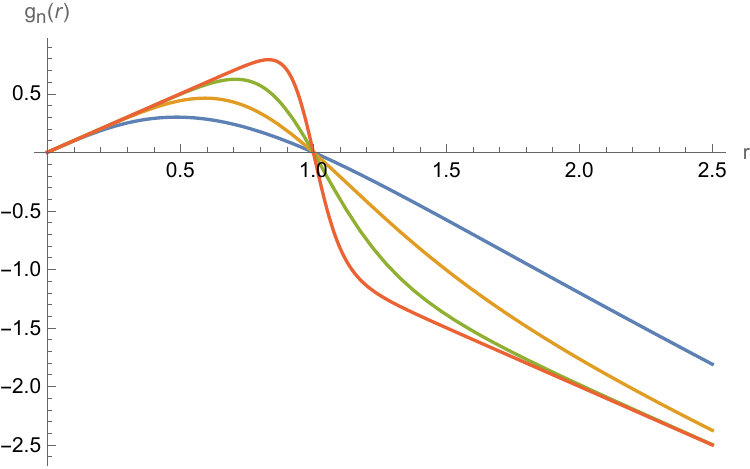}
        \caption{}
        \label{fig:gn_e2}
    \end{subfigure}
    \caption{\centering The plot of the function $g_n$ for $n=1,2,4,10$ in blue, yellow, green and red, respectively with (\subref{fig:gn_e06}) $\epsilon = 0.6$, (\subref{fig:gn_e08}) $\epsilon = 0.8$, (\subref{fig:gn_e1}) $\epsilon = 1$ and  (\subref{fig:gn_e2}) $\epsilon = 2$. The parameter $n$ has clear influence on shape of the function $g_n$, and therefore on qualitative behavior of solutions for certain inputs.}
    \label{fig:hillfunctions}
\end{figure}

The system to be analyzed emerges by inserting the expression for the steady states of weights $w_{1,2}$ into the first two equations of System \eqref{eq:all4bezd}:
\begin{equation}\label{eq:mainsystembezd}
    \begin{aligned}
        \tau_r \dot{r_1} &= -r_1 + \epsilon r_2 f(r_1r_2) +I_1, \\
        \tau_r \dot{r_2} &= -r_2 + \epsilon r_1 f(r_1r_2) +I_2 .
    \end{aligned}
\end{equation}
Solutions to System \eqref{eq:mainsystembezd} are also solutions to System \eqref{eq:all4bezd}, so they exist for all $t>0$ and are unique. We are interested in positive solutions. As a baseline we investigate a case with equal inputs, meaning that there is no stimulus, $I_1=I_2=I$. Checking for steady states we get:
\begin{equation*}
    \begin{aligned}
        r_1 &= w_2 r_2 + I,  \\
        r_2 &= w_1 r_1 + I ,
    \end{aligned}
\end{equation*}
and since $w_1=w_2$ we deduce that also $r_1=r_2 \eqqcolon r$.
% \aniaa{(Ja bym chyba usunęła/skróciła tę część) because equations for a steady state solution of System \eqref{eq:mainsystembezd} are 
% \begin{equation}\label{eq:steadystate}
%     \begin{alignedat}{3}
%         &\quad &&\qquad 0 &&= -r_1 + \epsilon r_2 f(r_1r_2) +I, \\
%         &\quad &&\qquad 0 &&= -r_2 + \epsilon r_1 f(r_1r_2) +I, \\
%         &\implies &&\qquad 0 &&= r_1 - r_2 + (\epsilon f(r_1r_2))(r_1-r_2), \\
%         &\implies &&\qquad r_1 &&= r_2 .
%     \end{alignedat}
% \end{equation}
% The last implication follows from the positivity of the coefficient $\epsilon$, the function $f$ and the variables $r_i$. }
To check the number of steady states we investigate the relationship between parameters $\epsilon, n$ and $I$ in a possible steady state. We introduce a function $g_n: \mathbb{R}_+ \to \mathbb{R}$ describing the relationship of $I$ and $r$ in a steady state.
%We introduce a function $g_n: \mathbb{R}_+ \to \mathbb{R}$, the condition for steady state being $I=g_n(r)$:
\begin{equation}\label{eq:funkcjagn}
    I = g_n(r) = r - \epsilon r f_n(r^2) = r - \epsilon \frac{r^{2n+1}}{1+r^{2n}}\, .
\end{equation}
The function $g_n$ has following properties dependent on parameters $\epsilon$ and $n$:
\begin{itemize}
    \item $\lim_{r \to 0} g_n(r) = 0$,
    \item for $\epsilon \in \Big(0, \frac{8n}{(2n+1)^2}\Big]$ the function $g_n$ is increasing and $\lim_{r \to \infty} g_n = +\infty $,
    \item for $\epsilon \in \Big(\frac{8n}{(2n+1)^2}, 1\Big)$ the function $g_n$ is increasing for $r<m_1$, decreasing for $r \in (m_1, m_2)$ and increasing for $r>m_2$ with $\lim_{r \to \infty} g_n = +\infty $,
    \item for $\epsilon = 1$ the function $g_n$ is increasing for $r<m_0$, decreasing for $r>m_0$ with $\lim_{r \to \infty} g_n = 0 $,
    \item for $\epsilon > 1$ the function $g_n$ is increasing for $r<m_2$, decreasing for $r>m_2$ with $\lim_{r \to \infty} g_n = - \infty $,
\end{itemize}
where $m_0, m_1, m_2$ are zeros of the derivative $\frac{d}{dr} g_n$ calculated -- and existing -- for $\epsilon = 1, \epsilon \in \Big( \frac{8n}{(2n+1)^2}, 1 \Big)$ and $\epsilon \in \Big( \frac{8n}{(2n+1)^2}, 1 \Big) \cup (1, +\infty)$ respectively. 
The number of steady states depends on a number of points at which the horizontal line $I$ crosses the graph of the function $g_n$, which leads to a conclusion that:
\begin{itemize}
    \item for $\epsilon \in \Big(0, \frac{8n}{(2n+1)^2}\Big]$ there exists one steady state for all $I>0$,
    \item for $\epsilon \in \Big(\frac{8n}{(2n+1)^2}, 1\Big)$ there exist
    \begin{itemize}
        \item one steady state for $I \in (0, g_n(m_2)) \cup (g_n(m_1), +\infty)$,
        \item two steady states for $I \in \{ g_n(m_1), g_n(m_2) \}$,
        \item three steady states for $I \in (g_n(m_2), g_n(m_1))$,
    \end{itemize}
    \item for $\epsilon = 1$ there exist
    \begin{itemize}
        \item one steady state for $I = g_n(m_0)$,
        \item two steady states for $I \in (0, g_n(m_0))$,
    \end{itemize}
    \item for $\epsilon \in (1, +\infty)$ there exist
    \begin{itemize}
        \item one steady state for $I = g_n(m_2)$,
        \item two steady states for $I \in (0, g_n(m_2))$.
    \end{itemize}
\end{itemize}
All solutions of the System \eqref{eq:mainsystembezd} with symmetric inputs lay on the straight line $r_1=r_2=r$. Therefore stability of the above steady states can be easily deduced from the derivative of firing rate $r$:
\begin{equation*}
    \dot{r} = I - r + \epsilon r f_n(r^2) = I - g_n(r)\, .
\end{equation*}
%\aniaa{(Dodałam $g_n$, zastanawiam się czy tego niżej nie dałoby się skrócić trochę)}
%\bartekk{ciut skrocilem}
It provides a simple geometrical way to check the stability of a steady state. If the horizontal line $I$ crosses the graph of the function $g_n$ at a point $(r_0, g_n(r_0))$ and $g_n$ is increasing in some neighborhood of $r_0$, then the derivative $\dot{r}$ is positive for smaller arguments $r<r_0$ and negative for greater arguments $r>r_0$. This means that such point $(r_0, r_0)$ is locally asymptotically stable, and in parallel when the graph of the function $g_n$ is crossed when it is decreasing, the point is unstable. If the monotonicity of the function differs on two sides of the point at which $I = g_n(r)$, the point is obviously unstable, but some solutions can still asymptotically tend to it. 
% \aniaa{Chyba jednak by się przydało wspomnieć, jakie są te stabilności u nas, bo potem na tym bazujesz w opóźnieniach, nie?}
In particular, the case with three equilibira is interesting. We denote values of firing rates at these points by $r_{I}, r_{II}, r_{III}$, $r_{I}< r_{II}< r_{III}$. Steady states $(r_I, r_I)$ and $(r_{III}, r_{III})$ are stable, and the state $(r_{II}, r_{II})$ is unstable. We later investigate the influence of the delay on stability of these equilibria. 
Noteworthy is also the fact that the number of steady states varies with differences in the values of the parameters $n, \epsilon$ and $I$. The Hill coefficient $n$ influences the function $g_n$ as well as the values $m_0, m_1, m_2$ and $\frac{8n}{(2n+1)^2}$. In particular, it can change the qualitative behavior of the system. This interplay is visible in the plots of $g_n$ for various $\epsilon$ and $n$ values, which are depicted in Figure \ref{fig:hillfunctions}.
% \bartek{obrazki tutaj - nie wiem ktore i ile, mozesz rzucic okiem w mojej mgr ( https://www.overleaf.com/project/6672b3395706a646b8b9796a ) na Figure 3, 4 i 5. ktores z tych bym tu dal, moze z Fig 4 (b) i a/d, z Fig 5 (b) np?}
% \ania{Pytanie jakie chcemy obrazki w sumie, bo ja u mnie w pracy mam tylko portrety fazowe pokazujące wpływ wielkości bodźca na 'podejmowaną decyzję'. I tam widać, że stany stacjonarne przestają istnieć dla innej wielkości bodźca dla różnych $n$. Może to warto dać do rozdziału z symulacjami? Jeśli chodzi o tę część, to mogę spróbować znaleźć kod i narysować jakieś inne, jeśli chcemy.}

\newpage
\subsection{Model with quasi-steady state approximation and non-zero delay}
% Two-equation model 

\noindent This model is a version of System \eqref{eq:mainsystembezd} with delay. It has previously been analyzed in \cite{forys2017impact} for $n=2$ with similar methods and results. The weights are in their steady states and the inputs are constant in time. The system takes the following form: 

% \noindent This model is a version of System \eqref{eq:all4bezd} \aniaa[]{A nie system 3?} with delay and has previously been analyzed in \cite{forys2017impact} for $n=2$ with similar methods and results. The weights $w_{1,2}$ are assumed to be in their steady states $w_{1,2} = \epsilon f_n(r_1r_2)$ at all times. As mentioned before, we also assume the inputs $I_{1,2}$ to be constant in time. \aniaa{These assumptions result in the System taking the following form:}
%This results in first two equations of System \eqref{eq:all4} taking the form
\begin{equation}\label{eq:mainsystem}
    \begin{aligned}
        \tau_r \dot{r_1} &= -r_1(t-\tau) + \epsilon r_2 f(r_1r_2) +I_1 ,\\
        \tau_r \dot{r_2} &= -r_2(t-\tau) + \epsilon r_1 f(r_1r_2) +I_2 .
    \end{aligned}
\end{equation}
The steady states of System \eqref{eq:mainsystem} are the same as those of zero-delay version of it, i.e. System \eqref{eq:mainsystembezd}. To investigate their stability we perform linearization at a generic point $(r_1, r_2)$, which yields a characteristic matrix $C(\lambda) = \lambda I - A - e^{-\lambda \tau} B$, where
\begin{equation*}
    \begin{aligned}
        A &= \frac{\epsilon}{\tau_r} 
        \begin{pmatrix}
         r_2^2 f'(r_1 r_2) & f(r_1 r_2) + r_1 r_2 f'(r_1 r_2) \\
        f(r_1 r_2) + r_1 r_2 f'(r_1 r_2) &  r_1^2 f'(r_1 r_2)
        \end{pmatrix}, \\[3pt]
        B &= \frac{1}{\tau_r} \begin{pmatrix}
            -1 & 0 \\
            0 & -1
        \end{pmatrix} .
    \end{aligned}
\end{equation*}
The determinant of $C$ then reads 
\begin{equation}\label{eq:determinant_almost}
\begin{aligned}
    \tau_r^2 \cdot \det C &= \big( \tau_r \lambda - \epsilon r_2^2 f'(r_1 r_2) + e^{-\lambda \tau} \big) \big( \tau_r \lambda - \epsilon r_1^2 f'(r_1 r_2) + e^{-\lambda \tau} \big) \\ &\quad - \epsilon^2 \big(f(r_1 r_2) + r_1 r_2 f'(r_1 r_2) \big)^2 .
\end{aligned}
\end{equation}
To allow for manageable analysis, we would like for the above expression to factorize. While it does not in the general case, it does at points on the line $r_1 = r_2 = r$. In the case of a symmetric system, where $I_1=I_2=I$, the system's steady states lie on this line and therefore we limit our symbolic analysis to this case.
%When the input is symmetric, meaning $I_1=I_2=I$, the steady states lie on that line, therefore we limit out symbolic analysis to this case. 
%When the equality $r_1=r_2$ holds, 
Now the expressions in the first two parentheses of \eqref{eq:determinant_almost} are equal, and the determinant is simplified to
\begin{equation*}
    \tau_r^2 \cdot \det C = \big( \tau_r \lambda - \epsilon f(r^2) - 2 \epsilon r^2 f'(r^2) + e^{-\lambda \tau} \big)
    \big( \tau_r \lambda + \epsilon f(r^2) + e^{-\lambda \tau} \big) .
\end{equation*}
From this, we can establish if and when the roots of the characteristic equation
\begin{equation}\label{eq:characteristic}
    0 = \underbrace{\big( \tau_r \lambda - \epsilon f(r^2) - 2 \epsilon r^2 f'(r^2) + e^{-\lambda \tau} \big)}_{W_1} \cdot
    \underbrace{\big( \tau_r \lambda + \epsilon f(r^2) + e^{-\lambda \tau} \big)}_{W_2}
\end{equation}
cross the imaginary axis. We are interested in the case where loss of stability is possible due to an increase in delay or a change in parameter $n$. We denote the smallest delay for which Equation \eqref{eq:characteristic} holds by $\tau_0 = \min \{\tau \ | \ W_1\cdot W_2 = 0\} $. It is equal to $ \tau_0 = \min \big\{\tau_0^1, \tau_0^2 \big\} $, for delays $\tau_0^i = \min \{ \tau \ |\ W_i = 0 \}$, $i=1,2$ being the smallest for which respective one of the factors $W_{1,2}$ is equal to zero. To find the values of $\tau_0^{1,2}$, we utilize the following theorem \cite{cooke1986zeroes}:
\begin{theorem}\label{tw:cooke}
    Let $P, Q: \mathbb{C} \to \mathbb{C}$ be analytic functions satisfying the conditions:
    \begin{enumerate}
        \item $P$ and $Q$ have no common imaginary zero,
        \item $\overline{P(iy)} = P(\overline{iy})$, $\overline{Q(iy)} = Q(\overline{iy})$ for $y \in \mathbb{R}$,
        \item $P(0) + Q(0) \neq 0$.
    \end{enumerate}
    Let also $F: \mathbb{R} \to \mathbb{R}$ be a real function defined as $F(y) = |P(iy)|^2 - |Q(iy)|^2$ and $s = \sgn F'(y)$ denote the sign of its derivative. Then for variables $\tau,\, z = iy$, $y \in \mathbb{R}_+$ satisfying the equation
    \begin{equation}\label{eq:twcooke}
        P(z) + Q(z)e^{-\tau z} = 0
    \end{equation}
    the following holds: if $z$ is a simple root and $s \neq 0$, then $y$ is a simple root of $F(y) = 0$ and the root $z(\tau)$ of Equation \eqref{eq:twcooke} crosses the imaginary axis in the direction given by $s$ as $\tau$ increases, meaning $s = \sgn \Big( \big( \frac{d}{d\tau} \operatorname{Re} z \big) \big|_{z=iy} \Big) $.
    %\begin{enumerate}
        %\item 
        %\item $y$ is a simple root of $F(y)=0$ if and only if $z=iy$ is a simple root of $|P(z)| - |Q(z)|=0$.
    %\end{enumerate}
\end{theorem}
We apply the above to both $W_1$ and $W_2$, first finding the points at which their complex zeros cross the imaginary axis. Because both expressions are of the form 
\begin{equation}\label{eq:transcendentalsimple}
    W_i(\lambda) = a \lambda + b + e^{-\lambda \tau}, 
\end{equation}
the function $F$ (using the notation from Theorem \ref{tw:cooke}) takes the form $F(y) = a^2y^2 + b^2 -1$ with its derivative equal to $F'(y) = 2a^2y$. Since $y$ is positive in the formulation of the theorem, the derivative $F'(y)$ is also positive, which means that the zeros of $W_{1,2}$ can only cross the imaginary axis from left to right as the delay $\tau$ increases. \par
\begin{figure}[h]
    \centering
    \begin{subfigure}[b]{0.45\linewidth}
        \centering
        \includegraphics[width=\linewidth]{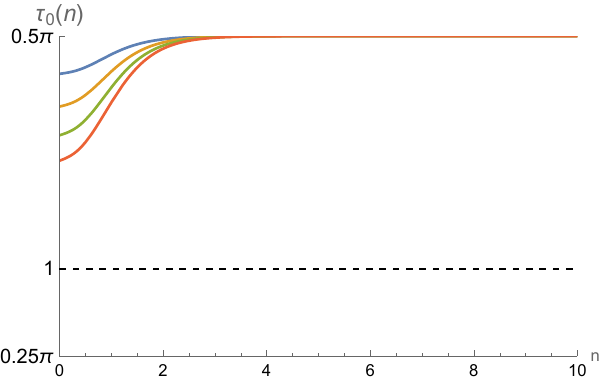}
        \caption{}
        \label{fig:tau_n_r03}
    \end{subfigure}
    \hfill
    \begin{subfigure}[b]{0.45\linewidth}
        \centering
        \includegraphics[width=\linewidth]{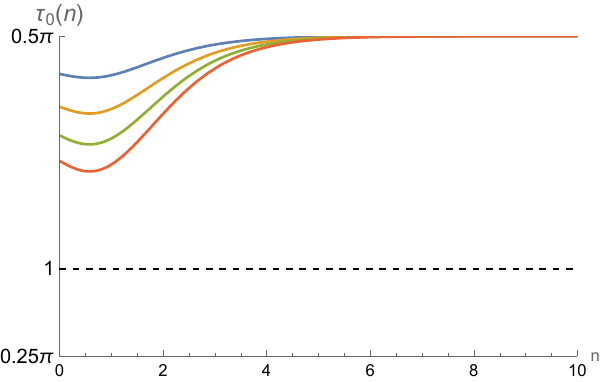}
        \caption{}
        \label{fig:tau_n_r05}
    \end{subfigure}
    \vfill
    \begin{subfigure}[b]{0.45\linewidth}
        \centering
        \includegraphics[width=\linewidth]{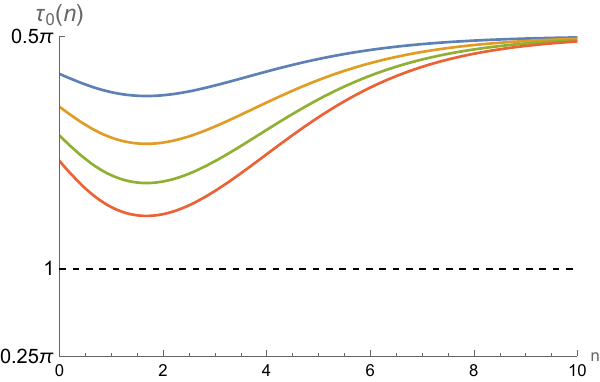}
        \caption{}
        \label{fig:tau_n_r07}
    \end{subfigure}
    \hfill
    \begin{subfigure}[b]{0.45\linewidth}
        \centering
        \includegraphics[width=\linewidth]{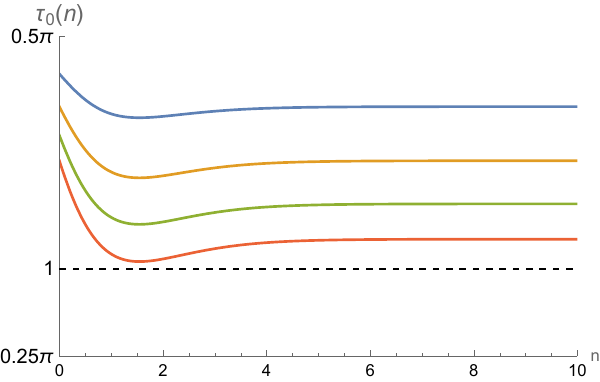}
        \caption{}
        \label{fig:tau_n_r2}
    \end{subfigure}
    \caption{\centering The plot of $\tau_0$ as a function of $n$ for $\epsilon=0.2, 0.4, 0.6, 0.8$ in blue, yellow, green and red, respectively with (\subref{fig:tau_n_r03}) $r = 0.3$, (\subref{fig:tau_n_r05}) $r = 0.5$, (\subref{fig:tau_n_r07}) $r = 0.7$ and  (\subref{fig:tau_n_r2}) $r = 2$. By choosing the value of $r$, the corresponding input $I$ is determined and there exist inputs for which each set of parameters correspond to a stable steady state for $\tau=0$. }
    \label{fig:taus}
\end{figure}

Solving the equalities $F(y)=0$ and $W_{1,2}(iy)=0$ for the general form of $W_{1,2}$ \eqref{eq:transcendentalsimple}, we obtain the values of $y$ and the delay $\tau$:
% for which they hold:
\begin{equation}\label{eq:arccos}
\begin{aligned}
    y &= \frac{\sqrt{1-b^2}}{|a|}, \\
    \bar{\tau}_k &= \frac{|a| (\arccos{(-b)} + 2k \pi)}{\sqrt{1-b^2}} \, , \ k \in \mathbb{Z}.
    \end{aligned}
\end{equation}
We are interested in positive values of $\bar{\tau}_k$, which implies nonnegative $k$. Note that the possible loss of stability occurs for $k=0$, as $\bar{\tau}_0$ is the smallest positive delay. For $W_{1,2}$, this leads to the following corresponding delay values:
\begin{equation*}
    \begin{aligned}
        \tau_0^1 &= \frac{ \tau_r \arccos{\big(\epsilon \cdot (f(r^2)+2r^2f'(r^2)) \big)} }{\sqrt{1-\big(\epsilon (f(r^2)+2r^2f'(r^2)) \big)^2}}, \\
        \tau_0^2 &= \frac{ \tau_r \arccos{\big(-\epsilon f(r^2) \big)} }{\sqrt{1-\big(\epsilon f(r^2) \big)^2}} .
    \end{aligned}
\end{equation*}
The necessary conditions for existence of the above delays (e.g. $\epsilon f(r^2) \subseteq (-1,1)$) are satisfied in the stable steady states $(r,r)$ due to basic properties of function $f$ and its derivative $f'$.
% Note that necessary inequalities $\epsilon f(r^2) \in (0,1) \subseteq (-1,1)$ and $\epsilon \cdot (f(r^2)+2f'(r^2)) \in (0,1) \subseteq (-1,1)$ are satisfied for stable steady states $(r,r)$. This is on one hand due to positivity of the function $f$ and its derivative $f'$, and on the other hand because of the form of the characteristic equation with zero delay:
% \begin{equation}\label{eq:characteristiczero}
%     0 = \big( \tau_r \lambda - \epsilon f(r^2) - 2 \epsilon r^2 f'(r^2) + 1 \big) \cdot
%     \big( \tau_r \lambda + \epsilon f(r^2) + 1 \big) .
% \end{equation}
% Stability of the steady state for the zero-delay case means that zeros of the above Equation \eqref{eq:characteristiczero} are negative, yielding 
% % \textcolor{red}{to jest chyba źle, bo to drugie wcale chyba z tego nie wynika??? już chyba ok}
% \begin{equation*}
%     \begin{aligned}
%         \epsilon \cdot (f(r^2)+2r^2f'(r^2)) &< 1, \\
%         \implies \epsilon f(r^2) &< 1 .
%     \end{aligned}
% \end{equation*}
The loss of stability happens for the smaller of the two delays $ \tau_0 = \min \big\{\tau_0^1, \tau_0^2 \big\} $. In order to know which one is smaller, we use the properties of the function $g,\ g:(-1,1) \to \mathbb{R}$ defined as $g(x) = \frac{\arccos{x}}{\sqrt{1-x^2}}$, since
\begin{equation}\label{eq:tau_g}
    \begin{aligned}
        \tau_0^1 &= \tau_r g\big(\epsilon \cdot (f(r^2)+2r^2f'(r^2)) \big), \\
        \tau_0^2 &= \tau_r g \big( -\epsilon f(r^2) \big) .
    \end{aligned}
\end{equation}
The function $g$ is monotonically decreasing and $\tau_0^1<\tau_0^2$.
%therefore comparing the expressions for $\tau_0^1$ and $\tau_0^2$ we get $\tau_0^1<\tau_0^2$. 
Moreover, $g(0) = \frac{\pi}{2}$, $\lim_{x\to1} g(x) = 1$ and $\big(\epsilon \cdot (f(r^2)+2r^2f'(r^2)) \big) > 0$, which implies $\tau_0^1 \in (\tau_r, \frac{\tau_r\pi}{2})$ regardless of the values of $r$, $\epsilon$ and $n$. 
\begin{figure}[H]
    \centering
    \includegraphics[width=0.5\linewidth]{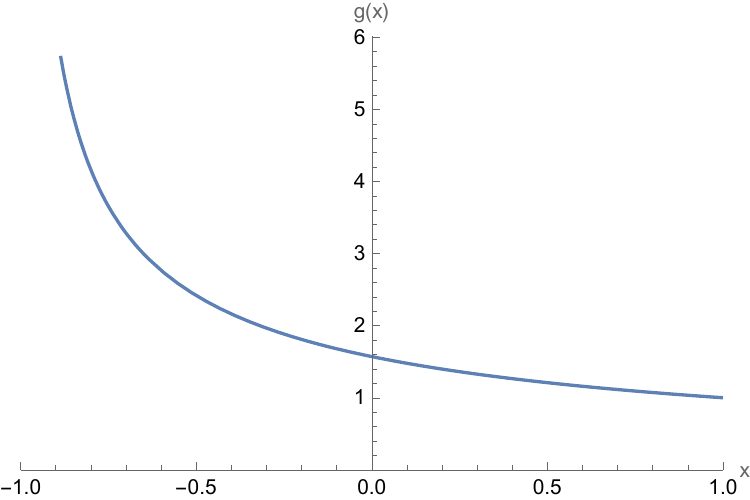}
    \caption{\centering The plot of the function $g(x) = \frac{\arccos{x}}{\sqrt{1-x^2}}$, where $\lim_{x \to -1}g(x) = \infty$ and $\lim_{x \to 1}g(x) = 1$.}
    \label{fig:arccos}
\end{figure} 
% \bartek{czy ok umiejscowienie obrazków?}
Assuming other parameters to be constant, the delay $\tau_0^1$ is decreasing with respect to $\epsilon$, while $\tau_0^2$ is increasing. This follows from the monotonicity of the function $g$ and the forms of $\tau_0^{1,2}$ in Equation \eqref{eq:tau_g} and also means that the distance between these two values is increasing with respect to $\epsilon$.
Examining the impact of the Hill coefficient $n$ or the firing rate $r$ on $\tau_0^1$ symbolically does not yield meaningful results and therefore we limit ourselves to numerical analysis, showing the impact of the coefficients $r, \epsilon$ and $n$.
% \aniaa{(A analiza wpływu $r$ albo $\epsilon$ jest możliwa? W sensie to zdanie nie do końca ma sens jeśli argumentujemy symulowanie wpływu wszystkich parametrów 'zachowaniem' jednego z nich)}. \bartekk{to jest monotoniczne ze względu na $\epsilon$, ze względu na $r$ prawie ale nie jest. myślisz że warto na to zwracać uwagę czy po prostu inaczej ubrac w slowa to co jest?}\aniaa{Wydaje mi się, że warto powiedzieć, co widać analitycznie, a co robimy tylko numerycznie i co te wykresy pokazują.} 
For this we assume $\tau_r = 1$, as it is only a scale factor. Results of the numerical simulations are shown in the Figure \ref{fig:taus}.
% \FloatBarrier

% \begin{figure}[h]
%     \centering
%     \begin{subfigure}[b]{0.48\linewidth}
%         \centering
%         \includegraphics[width=\linewidth]{Pictures/tau_n_e04_pi4.pdf}
%         \caption{}
%         \label{fig:tau_n_e04}
%     \end{subfigure}
%     \hfill
%     \begin{subfigure}[b]{0.48\linewidth}
%         \centering
%         \includegraphics[width=\linewidth]{Pictures/tau_n_e08_pi4.pdf}
%         \caption{}
%         \label{fig:tau_n_e08}
%     \end{subfigure}
%     \caption{\centering The plot of $\tau_0$ as a function of $n$ with for $r=0.2, 0.4, 0.6, 0.8$ in blue, yellow, green and red, respectively with (\subref{fig:tau_n_e04}) $\epsilon = 0.4$ and (\subref{fig:tau_n_e08}) $\epsilon = 0.8$. Similarly to Figure \ref{fig:taus}, all sets of parameters correspond to a stable steady state for $\tau=0$ and certain inputs $I$.}
    % \label{fig:hill}
% \end{figure}
% \FloatBarrier
% \bartek{na razie te obrazki mega rozjezdzaja plik, ale tym mysle ze na koniec sie najlepiej zajac. za to mozesz dac znac co uwazasz o ich umiejscowieniu w tekscie, czy ok? czy w ogole ktores bys wywalila?} 
% \ania{Ja bym dała chyba na koniec tego podrozdziału. I dopiała jakiś kawałek tekstu z opisem wniosków z wykresów. Coś jak tam pisaliśmy w wiadomościach, jak ja próbowałam to zrozumieć.}
\par
We have established that there exists such delay $\tau_0$ that a steady state stable for $\tau < \tau_0$ is unstable for $\tau > \tau_0$. At $\tau=\tau_0$ the Hopf bifurcation occurs, where a steady state transitions into periodic solutions. We state an appropriate theorem below, for which, we consider a system of equations with a real parameter $\mu$:
\begin{equation}\label{eq:hopf_param}
    \dot{x}(t) = f(\mu, x_t)\, .
\end{equation}
In this case specifically, the delay $\tau$ is the parameter we are interested in. We assume the function $f$ to be of class $C^1$, and for convenience that $0$ is a steady state solution of Equation \eqref{eq:hopf_param}, meaning $f(\mu, 0) = 0$ for all $\mu$.
\begin{theorem}\label{tw:bifurkacja}
    Let the characteristic equation of System \eqref{eq:hopf_param} for $\mu=0$ have an imaginary simple root $\lambda_0 = iy_0$ such, that: \begin{itemize}
        \item no other root $\lambda_i$ is an integer multiple of $\lambda_0$, meaning $\lambda_i \neq k \lambda_0$ for all $k \in \mathbb{Z}$,
        \item $\lambda_0$ crosses an imaginary axis with non-zero positive speed as parameter $\mu$ changes, meaning $ \frac{d}{d\mu} \operatorname{Re} (\lambda_0(\mu)) \Big\rvert_{\mu=0} > 0$.
    \end{itemize} 
    Then there exist positive constants $\epsilon_0, \mu_0, \delta, \beta > 0$, real functions $\mu$, $T$ of $\epsilon$ and a periodic solution $x^*(\epsilon)$ to Equation \eqref{eq:hopf_param} of period $T(\epsilon)$. Functions $\mu$, $T$ and $x^*$ are continuously differentiable for $|\epsilon| < \epsilon_0$. \par
    Moreover for $\mu, T$ with $|\mu|<\mu_0$, $\big|T - \frac{2\pi}{y_0}\big|<\delta$ every nonconstant $T$-periodic solution $x(t)$ of Equation \eqref{eq:hopf_param} satisfying $\max_t |x(t)| < \beta$ is of this form, up to a phase shift. 
    % \par
    % If $f$ is of class $C^5$, then also
    % \begin{equation*}
    % \mu(\epsilon) = \mu_1 \epsilon^2 + O(\epsilon^4)
    % \end{equation*}
    % and if all characteristic roots for $\mu = 0$ other than $ \pm \lambda_0 $ have strictly negative real parts, then $x^*$ is orbitally asymptotically stable if $\mu_1>0$ and orbitally unstable if $\mu_1<0$. The Hopf bifurcation is said to be \emph{supercritical} and \emph{subcritical} in these cases, respectively. 
    \end{theorem}
Note that the positive sign of the derivative $\frac{d}{d\mu} \operatorname{Re} (\lambda_0(\mu))$ is a normalization ensuring that for negative values of the parameter $\mu<0$ the root $\lambda_0(\mu)$ has negative real part, and for positive values of $\mu>0$ it has positive real part. The crucial assumption is that the derivative is non-zero. \par
% We can treat the delay $\tau$ as a parameter in the DDE\bartekk{ ogólnie to stwierdzenie wymaga pewnego uzasadnienia, ale chyba nie kiedy i tak pomijamy definicje itp}. 
With that, we see that a Hopf bifurcation occurs at $\tau = \tau_0$, as assumptions of Theorem \ref{tw:bifurkacja} are satisfied. 
% The function $f$ being of class $C^1$ is clear, as is the fact that no characteristic root is an integer multiple of $\lambda_0$, the one crossing the imaginary axis. 
The fact that the bifurcation occurs not at the value of the parameter being $0$ is purely cosmetic. 
The only not obvious assumption to be satisfied is the transversality condition, i.e. $ \frac{d}{d\tau} \operatorname{Re} (\lambda_0(\tau)) \Big\rvert_{\tau=\tau_0} > 0$. This is conveniently provided by the end part of Theorem \ref{tw:cooke}. 

\FloatBarrier 
\subsection{Three-equation model}

\noindent In this section we introduce and investigate a different simplified version of System \eqref{eq:all4}, which is the result of assuming the weights $w_{1,2}$ to be equal. This is the main novelty of the presented work comparing to previous results. From the original model we can obtain that
\begin{equation*}
    \begin{aligned}
        \frac{d}{dt}(w_1 - w_2)(t) &= - \frac{1}{\tau_w} (w_1 - w_2), \\
        \implies \quad (w_1 - w_2)(t) &= (w_1 - w_2)(0) \cdot e^{-\frac{t}{\tau_w}} .
    \end{aligned}
\end{equation*}
This shows the assumption $w_1 = w_2 = w$ to be justified -- it is the case if $w_1(0) = w_2(0)$. In this way this assumption resembles the quasi-steady-state approximation present in System \eqref{eq:mainsystem}, in which the weights are assumed to be in their steady states, while in the original System \eqref{eq:all4} they approach it exponentially with the exponent equal to $-\frac{t}{\tau_w}$. 
The resulting system reads:
\begin{equation}\label{eq:main3}
    \begin{aligned}
        \tau_r \dot{r}_1 &= -r_1(t-\tau) + wr_2+I_1, \\
        \tau_r \dot{r}_2 &= -r_2(t-\tau) + wr_1+I_2, \\
        \tau_w \dot{w} &= -w + \epsilon f(r_1r_2),
    \end{aligned}
\end{equation}
and once again the inputs $I_{1,2}$ are assumed to be constant in time. We begin with analysis of the zero-delay ODE system:
\begin{equation}\label{eq:main3bezd}
    \begin{aligned}
        \tau_r \dot{r}_1 &= -r_1 + wr_2+I_1, \\
        \tau_r \dot{r}_2 &= -r_2 + wr_1+I_2, \\
        \tau_w \dot{w} &= -w + \epsilon f(r_1r_2) .
    \end{aligned}
\end{equation}
% \bartek{tutaj nie wiem na ile się rozwodzić nad tym że stabilność stanów stacjonarnych jest taka sama jak dla 2 równań -- myślę że to można wziąc na wiarę w sumie?}
% \ania{Imo na wiarę git}
Its solutions can also be seen as solution of System \eqref{eq:all4bezd}, which means they exist for all $t>0$ and are unique. They are also positive for a positive initial condition. 
% \bartekk{może bez tego akapitu małego?} \aniaa[]{A to sądzę, że nie szkodzi, jak jest:)} 
\par
The steady states are the same as those of two-equation System \eqref{eq:mainsystembezd}, in a sense that the point $(r_1, r_2)$ is a steady state of that system if and only if the point $(r_1, r_2, \epsilon f(r_1r_2))$ is a steady state of System \eqref{eq:main3bezd} for the same inputs $I_{1,2}$. %For symmetric inputs 
In the symmetric case in which $I_1=I_2=I$, the number of steady states and its dependence on the parameters of the model are the same as well, since we can infer it from the same condition $I = g_n(r)$ with $g_n$ defined in Equation \eqref{eq:funkcjagn}. It seems that the stability of the steady state cannot be deduced only from the behavior of the function $g_n$, because of the weight $w$ not being in equilibrium at all times. However, it is easy to check through basic linearization that the stability's dependence on parameters is exactly the same for Systems \eqref{eq:main3} and \eqref{eq:mainsystem}. This serves as a convenient starting point for the non-zero delay version. 
\par
To investigate the steady states' stability for System \eqref{eq:main3}, we linearize it at a generic point $(r_1, r_2, w)$, obtaining the characteristic matrix $C = \lambda I - A - e^{-\lambda \tau} B$, where
\begin{equation*}
    \begin{aligned}
        A &= \begin{pmatrix}
0 & \frac{w}{\tau_r} & \frac{r_2}{\tau_r} \\
\frac{w}{\tau_r} & 0 & \frac{r_1}{\tau_r} \\
\epsilon \frac{r_2f'(r_1 r_2)}{\tau_w} & \epsilon \frac{r_1f'(r_1 r_2)}{\tau_w} & -\frac{1}{\tau_w} \\ 
\end{pmatrix},  \\
B &= \begin{pmatrix}
-\frac{1}{\tau_r} & 0 & 0 \\
0 & -\frac{1}{\tau_r} & 0 \\
0 & 0 & 0
\end{pmatrix} .
    \end{aligned}
\end{equation*}
The determinant of $C$ then reads
\begin{equation*}
\begin{split}
    \tau_r^2 \tau_w \cdot \det C = -w^2-2(\epsilon f'(r_1 r_2) r_1 r_2 w)+\tau_r^2 \tau_w \lambda^3 + \tau_r^2 \lambda^2-\epsilon f'(r_1 r_2) r_1^2 \tau_r \lambda \\ -\epsilon f'(r_1 r_2) r_2^2 \tau_r \lambda - \tau_w w^2 \lambda + 
    \tau_w \lambda e^{-2t\lambda}+e^{-2t\lambda}-\epsilon f'(r_1 r_2) r_1^2 e^{-\tau \lambda} \\-\epsilon  f'(r_1 r_2) r_2^2 e^{-\tau \lambda} +2 (\tau_r \tau_w \lambda^2 e^{-\tau \lambda})+2 (\tau_r \lambda e^{-\tau \lambda}).
\end{split}
\end{equation*}
As it does not factorize in the general case, we again limit our analysis to the case $I_1 = I_2 = I$, which implies that in a steady state $r_1=r_2$. Then the characteristic function reads:
\begin{equation*}
\begin{aligned}
    \tau_r^2 \tau_w  \det C &= -w^2 - 2(\epsilon f'(r^2) r^2 w)+ \tau_r^2 \tau_w \lambda^3+ \tau_r^2 \lambda^2- 2 (\epsilon f'(r^2) r^2 \tau_r \lambda) \\ 
    &\quad - \tau_w w^2 \lambda +\tau_w \lambda e^{-2\tau \lambda}+ e^{-2 \tau \lambda}- 2 (\epsilon f'(r^2) r^2 e^{-\tau \lambda}) \\ 
    &\quad + 2 (\tau_r \tau_w \lambda^2 e^{-\tau \lambda})+ 2(\tau_r \lambda e^{-\tau \lambda}) \\
    &=
    \big( \tau_r \tau_w \lambda^2 + (\tau_r - \tau_w w)\lambda - 2 \epsilon f'(r^2) r^2 - w + (\tau_w \lambda + 1) e^{-\tau \lambda} \big) \\
    &\quad \cdot (\tau_r \lambda + w + e^{-\tau \lambda}).
\end{aligned}
\end{equation*}
From this, we establish if and when the roots of the characteristic equation
\begin{equation}\label{eq:characteristic3}
\begin{aligned}
    0 &= \underbrace{(\tau_r \lambda + w + e^{-\tau \lambda})}_{W_1(\lambda)} \\ 
    &\cdot \underbrace{
    \big( \tau_r \tau_w \lambda^2 + (\tau_r - \tau_w w)\lambda - 2 \epsilon f'(r^2) r^2 - w + (\tau_w \lambda + 1) e^{-\tau \lambda} \big)}_{W_2(\lambda)}    
\end{aligned}
\end{equation}
cross the imaginary axis. Note that for $\tau_w=0$ the above simplifies to the characteristic equation of the quasi-steady state approximation model, as seen in Equation \eqref{eq:characteristic}. \par
Firstly, we use the previous results to investigate when the root of the function $W_1$ crosses the imaginary axis, as it is of the form presented in Equation \eqref{eq:transcendentalsimple}. Denote the smallest delay for which this is the case by $\tau_0^1 = \min \{ \tau \ |\ W_1 = 0 \}$. From Equation \eqref{eq:arccos} with parameter $k=0$ we obtain:
\begin{equation}\label{eq:tau1}
\begin{aligned}
        \tau_0^1 &= \frac{ \tau_r \arccos{(-w )} }{\sqrt{1-w^2}}
        = \frac{ \tau_r \arccos{\big(-\epsilon f(r^2) \big)} }{\sqrt{1-\big(\epsilon f(r^2) \big)^2}},
\end{aligned}
\end{equation}
where the second equality holds because in a steady state $w = \epsilon f(r^2)$. The point at which the root crosses the imaginary axis has coordinates $(0, iy_1)$, where: 
\begin{equation}\label{eq:y1}
    y_1 = \frac{\sqrt{1-(\epsilon f(r^2))^2}}{\tau_r}.
\end{equation}
\par We now investigate the roots of the function $W_2(\lambda)$ -- however, it takes the form different to $W_1$, having one-degree higher polynomials, i.e.
\begin{equation}\label{eq:transcendentalhard}
    W_2(\lambda) = \underbrace{a \lambda^2 + b \lambda + c}_{P(\lambda)} + \underbrace{(\alpha \lambda + \beta)}_{Q(\lambda)} e^{-\tau \lambda} .
\end{equation}
% \bartek{poniższy fragment jest do sprawdzenia, ale w miare ok}
Although this is a generalization of the function from Equation \eqref{eq:transcendentalsimple}, we investigate it separately due to the possible qualitative differences in behavior. We again use Theorem \ref{tw:cooke} and its notation, with function $F$ that reads
\begin{equation*}
\begin{aligned}
    F(y) 
    &= a^2y^4 + (b^2-2ac - \alpha^2 )y^2 + c^2 - \beta^2 .
    \end{aligned}
\end{equation*}
We substitute $z = y^2$, $z>0$, having $y=\sqrt{z}$ and introducing an auxiliary function $\bar{F}$: 
\begin{equation}\label{eq:FibarF}
\begin{aligned}
    F(\sqrt{z}) = \bar{F}(z) &= a^2z^2 + (b^2-2ac - \alpha^2 )z + c^2 - \beta^2, \\
    \Delta &= (b^2-2ac - \alpha^2 )^2 - 4a^2(c^2-\beta^2), \\
    z_{\pm} &= \frac{2ac + \alpha^2 - b^2 \pm \sqrt{\Delta}}{2a^2}, \\
    y_{\pm} &= \bigg( \frac{2ac + \alpha^2 - b^2 \pm \sqrt{\Delta}}{2a^2} \bigg)^{\frac{1}{2}}.
\end{aligned}
\end{equation}
We are only interested in $y>0$, as stated in Theorem \ref{tw:cooke}. Sign of the determinant $\Delta$ depends on the parameters. If it is negative, then the function $W_2$ has no purely imaginary roots and the change of stability is not possible. 
If it is nonnegative, then the signs of $z_-, z_+$ need to be investigated and for each positive one of them there exists a purely imaginary root of the function $W_2$, $iy_-$ or $iy_+$ respectively. The derivative of the function $F$ takes the form
\begin{equation}\label{eq:pochodnaFy}
        F'(y) = 2y (2a^2y^2 + b^2-2ac - \alpha^2 ),
\end{equation}
with its sign dependent on the parameters.
For the specific form of $W_2$, the function $\bar{F}$ is
\begin{equation*}
\begin{aligned}
        \bar{F}(z) &= \tau_r^2 \tau_w^2 z^2 + \big( (\tau_r - \tau_w w)^2 - 2 \tau_r \tau_w (-2 \epsilon r^2 f'(r^2) - w) - \tau_w^2 \big) z 
        \\ &+ \big(-2 \epsilon r^2 f'(r^2) - w \big)^2 - 1 \\
        &= \tau_r^2 \tau_w^2 z^2 + \big( 4 r^2 \epsilon \tau_r \tau_w f'(r^2) + \tau_r^2 + (w^2 - 1) \tau_w^2 \big) z \\&+ \big(2 r^2 \epsilon f'(r^2) + w\big)^2 - 1 .
\end{aligned}
\end{equation*}
The discriminant $\Delta$ reads 
% \bartekk{musze tu porozdzielac na kilka linii}
\begin{equation}\label{eq:delta}
\begin{aligned}
    \Delta &= \big( 4 r^2 \epsilon \tau_r \tau_w f'(r^2) + \tau_r^2 + (w^2 - 1) \tau_w^2 \big)^2 \\&- 4 \tau_r^2 \tau_w^2 \big((2 r^2 \epsilon f'(r^2) + w)^2 - 1 \big),
\end{aligned}
\end{equation}
and its sign is not clear in the general case. From the analysis of the zero-delay case it is easy to infer the sign of the  expression $(2 \epsilon r^2 f'(r^2) + w)^2 - 1$ in the steady state, which depends on its stability for $\tau=0$. If it is stable, then $(2 \epsilon r^2 f'(r^2) + w)^2 - 1 < 0$ and the coefficient $\Delta$ is positive, meaning there exist two roots $z_-, z_+$ of the function $\bar{F}$. We do not need to find the exact expression for each of them to know their signs, because $\bar{F}(0) = (2 \epsilon r^2 f'(r^2) + w)^2 - 1 < 0$, which means that exactly one of them is positive, i.e. $z_+$. \par
% \bartek{do zmiany} 
From the forms of $y_+$ as seen in \eqref{eq:FibarF} and $F'(y)$ as seen in \eqref{eq:pochodnaFy} we can see that $F'(y_+) > 0$, which means that the complex zeros of the characteristic function can only cross the imaginary axis from left to right. Consequently, as the delay $\tau$ increases and reaches a critical value, a stable steady state loses stability. 

\par We now need to answer a question whether an unstable steady state might become stable through an increase in delay, which would be an unexpected behavior, different from that of solutions to System \eqref{eq:mainsystem}.
% Simultaneously, an unstable (for $\tau=0$) steady state cannot gain it due to an increase in the delay. 
For an unstable steady state $\bar{F}(0) \geq 0$ holds. From the analysis of the zero-delay case one can easily find that if a steady state is unstable, it is a saddle and there exists one positive and two negative zeros of the characteristic polynomial. The characteristic function takes the form $W(\lambda) = P(\lambda) + Q(\lambda) e^{-\tau \lambda} = \tau_r^2\tau_w \lambda^3 + \dots$, where $\deg P > \deg Q$, which means that $\lim_{\lambda \to \infty} W(\lambda) = \infty$. For the zero-delay case we have $W(\lambda) = P(\lambda) + Q(\lambda)$, and the limit of $W$ in infinity and the existence of exactly one positive zero mean that $W(0) < 0$. Note that for the non-zero delay this still holds, because $W(0) = P(0) + Q(0)$ independent of the delay. Consequently, treating $W$ as a real function we use the intermediate value theorem and deduce from $W(0) < 0$ and $\lim_{\lambda \to \infty} W(\lambda) = \infty$ that there exists at least one positive zero of the characteristic function for the case with the delay. This means that a steady state unstable for $\tau=0$ cannot gain stability for $\tau>0$. \par
\begin{figure}[H]
    \centering
    \begin{subfigure}[b]{0.45\linewidth}
        \centering
        \includegraphics[width=\linewidth]{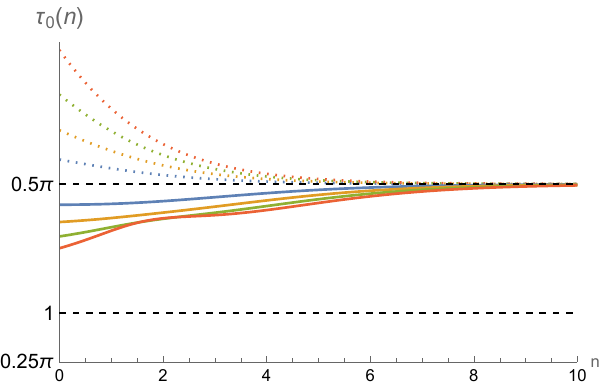}
        \caption{}
        \label{fig:tau_n_tw03_r07}
    \end{subfigure}
    \hfill
    \begin{subfigure}[b]{0.45\linewidth}
        \centering
        \includegraphics[width=\linewidth]{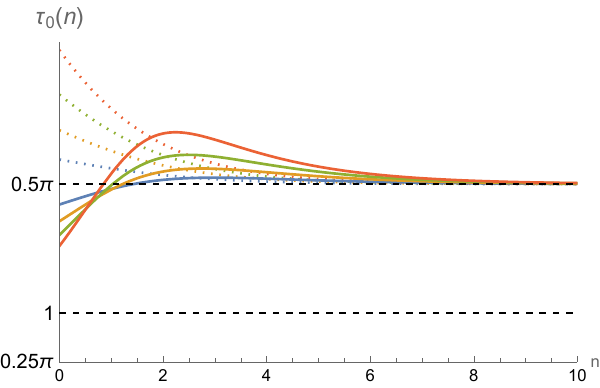}
        \caption{}
        \label{fig:tau_n_tw08_r07}
    \end{subfigure}
    \vfill
    \begin{subfigure}[b]{0.45\linewidth}
        \centering
        \includegraphics[width=\linewidth]{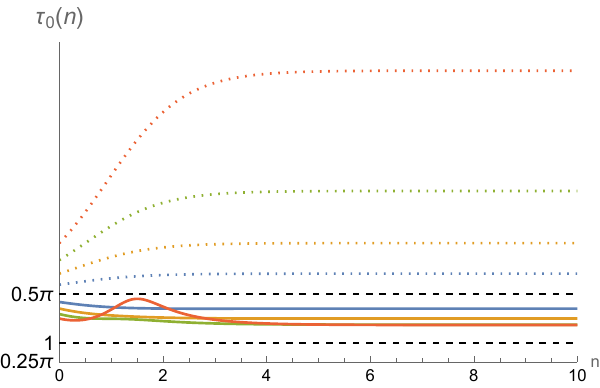}
        \caption{}
        \label{fig:tau_n_tw03_r2}
    \end{subfigure}
    \hfill
    \begin{subfigure}[b]{0.45\linewidth}
        \centering
        \includegraphics[width=\linewidth]{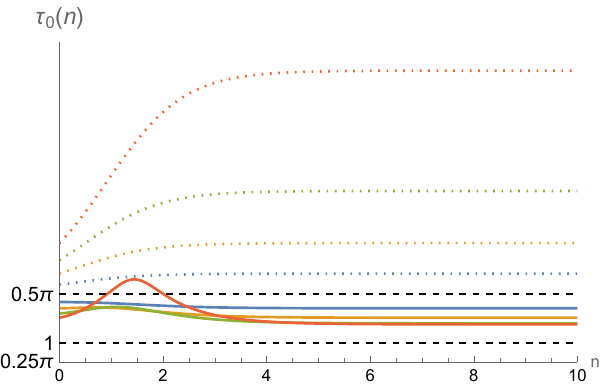}
        \caption{}
        \label{fig:tau_n_tw08_r2}
    \end{subfigure}
    \caption{\centering The plots of $\tau_0^2$ (solid lines) and $\tau_0^1$ (dotted lines) as functions of $n$ for $\epsilon=0.2, 0.4, 0.6, 0.8$ in blue, yellow, green and red, respectively with (\subref{fig:tau_n_tw03_r07}) $\tau_w = 0.3$, $r = 0.7$, (\subref{fig:tau_n_tw08_r07}) $\tau_w = 0.8$, $r = 0.7$, (\subref{fig:tau_n_tw03_r2}) $\tau_w = 0.3$, $r = 2$ and  (\subref{fig:tau_n_tw08_r2}) $\tau_w = 0.8$, $r = 2$. By choosing the value of $r$, the corresponding input $I$ is determined and there exist inputs for which each set of parameters correspond to a stable steady state for $\tau=0$. }
    \label{fig:taus_3eq}
\end{figure}

We investigate the situation in which a steady state stable for $\tau=0$ loses its stability with increase in the delay. We have established that the root of the function $W_2$ crosses the imaginary axis from the negative side to the positive at the point $(0, iy_+)$, where
\begin{equation}\label{eq:y+}
\begin{aligned}
    y_+ &= \bigg( \frac{2ac + \alpha^2 - b^2 + \sqrt{\Delta}}{2a^2} \bigg)^{\frac{1}{2}} \\
    &= \bigg( \frac{- \big( 4 r^2 \epsilon \tau_r \tau_w f'(r^2) + \tau_r^2 + \big((\epsilon f(r^2))^2 - 1 \big) \tau_w^2 \big) + \sqrt{\Delta}}{2 \tau_r^2 \tau_w^2} \bigg)^{\frac{1}{2}},
\end{aligned}
\end{equation}
and the coefficient $\Delta$ is described by Equation \eqref{eq:delta} with $w = \epsilon f_n(r^2)$. We denote the smallest corresponding delay by $\tau_0^2 = \min\{ \tau \ | \ W_2 = 0 \}$, and from the equality $W_2(iy_+)=0$ with the general form \eqref{eq:transcendentalhard} of $W_2$ we get the system of equations:
\begin{equation*}
        \left\{ \ 
        \begin{aligned}
          \beta \cos(y_+\tau) +\alpha y_+ \sin(y_+\tau) &=  ay_+^2 -c, \\
          \alpha y_+ \cos(y_+\tau) - \beta \sin(y_+\tau) &= -b .
        \end{aligned} 
        \right.
\end{equation*}
This results in
\begin{equation}\label{eq:secondcasedelay}
    \begin{aligned}
        \cos (y_+\tau) &= \frac{a \beta y_+^2 - \alpha b y_+ - c \beta}{\alpha^2 y_+^2 + \beta^2}, \\
        \sin (y_+\tau) &= \frac{a \alpha y_+^3- \alpha c y_+ +b \beta}{\alpha^2 y_+^2 +\beta^2}.
    \end{aligned}
\end{equation}
To find the corresponding delay $\tau$ we need to know the sign of the expression for $\sin (y_+\tau)$ -- substituting the specific parameters in the expression one can obtain, with some assumptions, that it is positive. One example of such sufficient assumption is $\tau_w \leq \tau_r$, which is satisfied for this model. This means that there exists such delay $\tau$ that \eqref{eq:secondcasedelay} is satisfied, and it takes the following form:
\begin{equation}\label{eq:actualsecondcasedelay}
    \bar{\tau}_k = \frac{1}{y} \bigg( \arccos \Big( \frac{a \beta y^2 - \alpha b y - c \beta}{\alpha^2 y^2 + \beta^2} \Big) + 2k\pi \bigg)
\end{equation}
for $k \in \mathbb{N}$. Only $\bar{\tau}_0$ is of interest as the smallest positive delay, the one for which change of stability occurs. For the specific form of $W_2$ \eqref{eq:characteristic3} the delay \eqref{eq:actualsecondcasedelay} with $k=0$ becomes
\begin{equation}\label{eq:tau2}
    \begin{aligned}
        \tau_0^2 &= \frac{1}{y_+} \arccos \Big( \frac{\tau_r \tau_w  y_+^2 - \tau_w (\tau_r - \tau_w \epsilon f(r^2) ) y_+ + 2 r^2 \epsilon f'(r^2) + \epsilon f(r^2)}{\tau_w^2 y_+^2 + 1} \Big).
    \end{aligned}
\end{equation}
The steady state which is stable for $\tau=0$ loses its stability for $\tau = \tau_0 = \min \{ \tau_0^1, \tau_0^2 \}$ with $\tau_0^1$ described by the equation \eqref{eq:tau1}. It is unclear if one of the possible critical values of $\tau$ is smaller regardless of the parameters, which was the case for the two-equation model. However, due to the closed-form expression for both of them, the smaller value can be determined for any specific set of coefficients. Figure \ref{fig:taus_3eq} shows plots of the critical delays generated numerically for different parameter sets. For ease of comparison with Figure \ref{fig:taus}, we assume $\tau_r=1$.
% \bartekk{For numerically generated plots we assume $\tau_r=1$ for easy comparison to Figure \ref{fig:taus}}. 
% \bartek{tu obrazki opoznien ktore musze zrobic jeszcze}
% \FloatBarrier
% \par
The delays shown in Figure \ref{fig:taus_3eq} exhibit several properties of interest. Firstly, it is worth noting that, contrary to Figure \ref{fig:taus}, $\tau_0^2$ is not necessarily in the interval $(\tau_r, \frac{\tau_r\pi}{2})$. The other delay $\tau_0^1$ is in fact always larger than $\frac{\tau_r\pi}{2}$. Secondly, while the delay $\tau_0^1$ and the delays in Figure \ref{fig:taus} are monotonous with respect to $\epsilon$, the delay $\tau_0^2$ does not have such property. These two properties can be deduced directly from the forms of the delays in Equations \eqref{eq:tau1} and \eqref{eq:tau2}. Lastly, in Figure \ref{fig:taus_3eq}(\subref{fig:tau_n_tw08_r07}) one can see that there exist parameters for which $\tau_0^1 <\tau_0^2$ -- a behavior different from that in the two-equation System \eqref{eq:mainsystem}, where the delay corresponding to one factor of the characteristic function was smaller than the other for any set of parameters.
%\bartekk{The delays shown in Figure \ref{fig:taus_3eq} exhibit several properties of interest. Firstly, it is worth noting that contrary to Figure \ref{fig:taus}, $\tau_0^2$ is not necessarily in the interval $(\tau_r, \frac{\tau_r\pi}{2})$. The other delay $\tau_0^1$ is in fact always larger than $\frac{\tau_r\pi}{2}$. Secondly, while the delay $\tau_0^1$ and the delays in Figure \ref{fig:taus} are monotonous with respect to $\epsilon$, the delay $\tau_0^2$ does not have such property. These two properties can be deduced directly from the delays' forms in Equations \eqref{eq:tau1} and \eqref{eq:tau2}. Lastly, in Figure \ref{fig:taus_3eq}(\subref{fig:tau_n_tw08_r07}) one can see that there exist parameters for which $\tau_0^1 <\tau_0^2$, a behavior different from that of delays for two-equation System \eqref{eq:mainsystem}, in which the delay corresponding to one factor of the characteristic function was smaller than the other for any set of parameters.}
With the delay $\tau$ as a parameter, we check if the assumptions of Theorem \ref{tw:bifurkacja} are satisfied at $\tau = \tau_0$. For the same clear reasons as in the two-equation model \eqref{eq:mainsystem}, all but one of them are. If the delays $\tau_0^1$, $\tau_0^2$ turn out to be equal, one needs to check if one of corresponding points at which the two roots of the characteristic equation cross the imaginary axis is an integer multiple of the other one. Specifically, the points corresponding to the delays are $(0,iy_1), (0,iy_+)$ respectively, where $y_1$ and $y_+$ are described by Equations \eqref{eq:y1} and \eqref{eq:y+}. If one of them is an integer multiple of the other one, Theorem \ref{tw:bifurkacja} cannot be applied. If this is not the case, then for the smaller one of the delays $\tau_0^1, \tau_0^2$ the stability is lost and the Hopf bifurcation occurs. 
\FloatBarrier
\section{Numerical simulations}

\noindent In this section Systems \eqref{eq:mainsystem} and \eqref{eq:main3} are investigated using numerical methods. 
The simulations were performed using Wolfram Mathematica 13.0 and the built-in delay differential equation solving tools. We compare the behavior of solutions of the two systems for different sets of nine coefficients:
\begin{itemize}
    \item $\epsilon, \tau, n$ -- the maximal synapse capacity, the delay and the Hill coefficient, respectively,
    \item $I_1, I_2$ -- the constant inputs for each neuron population,
    \item $\tau_r, \tau_w$ -- time scales for the firing rates and the weight, respectively,
    \item $r_1^0, r_2^0$ -- the constant value of each firing rate on the interval $[-\tau, 0]$, serving as the initial condition.
\end{itemize}
\begin{table}[H]
\centering
\small
\begin{tabular}{@{\extracolsep{\fill}} |c|c|c|c|c|c|c|c|c|c|}
\hline
\backslashbox{set}{coef.} & $n$ & $\epsilon$ & $\tau$ & $I_1$ & $I_2$ & $\tau_r$ & $\tau_w$ & $r_1^0$ & $r_2^0$ \\ \hline
% $1$  & 2   & 0.6        & 0      & 0.4   & 0.6   & 1        & 0.2      & ---       & ---       \\\hline
% $2$  & 2   & 0.6        & 0.6    & 0.4   & 0.6   & 1        & 0.5      & 0       & 1       \\\hline
% $3$ 
$1$ & 2   & 0.6        & 0.6    & 0.6   & 0.7   & 1        & 0.5      & 0       & 1       \\\hline
% $4$ 
$2$ & 2   & 0.6        & 0.9    & 0.6   & 0.7   & 1        & 0.5      & 0       & 1       \\\hline
% $5$  & 2   & 0.8        & 0.6    & 0.6   & 0.7   & 1        & 0.5      & 0       & 1       \\\hline
% $7$  & 2   & 0.4        & 1.2    & 0.6   & 0.7   & 1        & 0.5      & 0       & 1       \\\hline
% $8$  
$3$ & 2   & 0.6        & 1.2    & 0.6   & 0.7   & 1        & 0.5      & 0       & 1       \\\hline
% $9$
$4$ & 4   & 0.6        & 1.2    & 0.6   & 0.7   & 1        & 0.5      & 0       & 1       \\\hline
% $6$
$5$ & 2   & 0.72       & 0.8    & 0.6   & 0.7   & 1        & 0.5      & 0       & 1       \\\hline
% $10$
$6$ & 4   & 0.72       & 0.8    & 0.6   & 0.7   & 1        & 0.5      & 0       & 1       \\\hline
% $11$ 
$7$ & 1   & 0.72       & 0.8    & 0.6   & 0.7   & 1        & 0.5      & 0       & 1       \\\hline
% $12$ & 2   & 1.5        & 0.6    & 0.4   & 0.6   & 1        & 0.5      & 0       & 1       \\\hline
% $13$ & 2   & 1.65       & 0.6    & 0.4   & 0.6   & 1        & 0.5      & 0       & 1       \\\hline
% $14$ & 2   & 0.6        & 0.6    & 0.6   & 0.7   & 2        & 0.5      & 0       & 1       \\\hline
% $15$ & 2   & 0.6        & 0.6    & 0.6   & 0.7   & 0.5        & 0.5      & 0       & 1       \\\hline
% $16$  & 2   & 0.6        & 0.6    & 0.6   & 0.7   & 1        & 5      & 0       & 1       \\\hline
\end{tabular}
\caption{The sets of coefficients used in plots.}
\label{table:1}
\end{table}
The weight was assumed to be in its steady state during the time $[-\tau, 0]$, meaning $w=\epsilon f(r_1r_2)$, equal to $0$ for the chosen coefficient sets.
The different sets of the coefficients, presented in Table \ref{table:1}, were considered to show the impact of certain coefficients. 
The behavior of the solutions of Systems \eqref{eq:mainsystem} and \eqref{eq:main3} was compared for each set with $6$ various plots. The plots (a), (c) and (e) on the left are for System \eqref{eq:main3} and the plots (b), (d) and (f) on the right are for System \eqref{eq:mainsystem}. The plots (a) and (b) show the values of the firing rates $r_1$ and $r_2$ in time. In the plot (a), the blue line is the variable $r_1$ and the yellow line is the variable $r_2$. In the plot (b), the green line is the variable $r_1$ and the red line is the variable $r_2$. 
The plots (c) and (d) show the variables $r_{1,2}$ in time $t$ on the $x$ axis and in time $t-\tau$ on the $y$ axis. When the solutions fluctuate, this kind of plot is useful to see how significant the fluctuations are. The colors used are the same as in the plots (a) and (b). 
The plot (e) shows the weight variable $w$ in time, a~solution to System \eqref{eq:main3}. The dashed line is the value $\epsilon f(r_1r_2)$, for which $\dot{w}=0$. In the plot (f) there is the steady state  value $w = \epsilon f(r_1r_2)$ for System \eqref{eq:mainsystem}.
\vfill
\begin{figure}[H]
    \centering
    \begin{subfigure}[b]{0.45\linewidth}
        \centering
        \includegraphics[width=\linewidth]{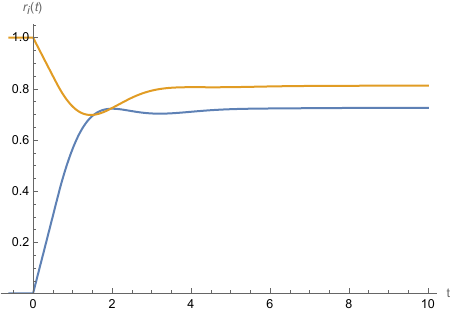}
        \caption{}
        % \label{fig:tau_n_r03}
    \end{subfigure}
    \hfill
    \begin{subfigure}[b]{0.45\linewidth}
        \centering
        \includegraphics[width=\linewidth]{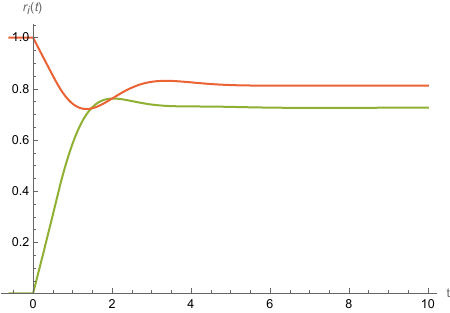}
        \caption{}
        % \label{fig:tau_n_r05}
    \end{subfigure}
    \vfill
    \begin{subfigure}[b]{0.45\linewidth}
        \centering
        \includegraphics[width=\linewidth]{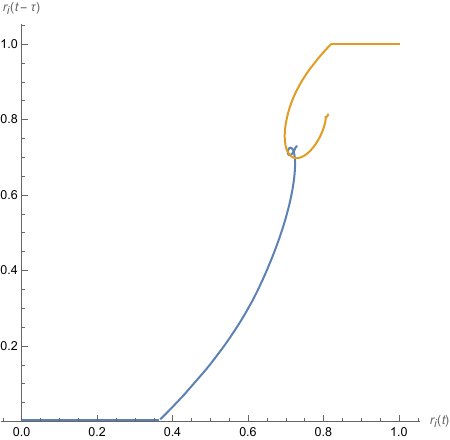}
        \caption{}
        % \label{fig:tau_n_r03}
    \end{subfigure}
    \hfill
    \begin{subfigure}[b]{0.45\linewidth}
        \centering
        \includegraphics[width=\linewidth]{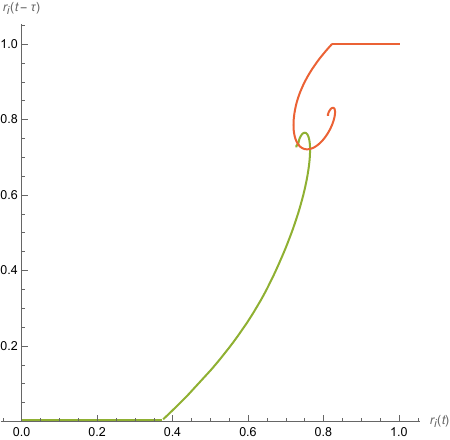}
        \caption{}
        % \label{fig:tau_n_r05}
    \end{subfigure}
    \vfill
    \begin{subfigure}[b]{0.45\linewidth}
        \centering
        \includegraphics[width=\linewidth]{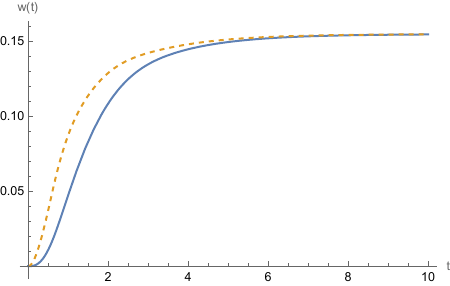}
        \caption{}
        % \label{fig:tau_n_r07}
    \end{subfigure}
    \hfill
    \begin{subfigure}[b]{0.45\linewidth}
        \centering
        \includegraphics[width=\linewidth]{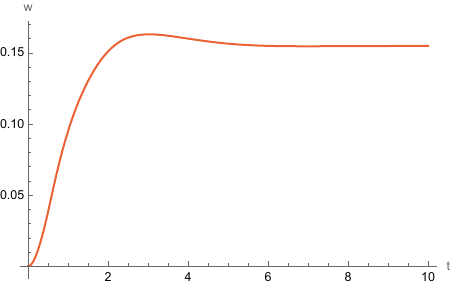}
        \caption{}
        % \label{fig:tau_n_r2}
    \end{subfigure}
    \caption{\centering The plots for the coefficient set $1$ from Table \ref{table:1}.}
    \label{fig:3set}
\end{figure}
The plots in Figure \ref{fig:3set} show the similarity of solutions of the two systems. However, there are some minor differences, for example the non-monotonicity of the weight $w$ in Fig. \ref{fig:3set}(f). Worth mentioning is also the fact that for a brief period of time $r_2<r_1$ for both systems, in spite the fact that for the rate $r_2$ both the input and the initial condition are greater.
\begin{figure}[H]
    \centering
    \begin{subfigure}[b]{0.45\linewidth}
        \centering
        \includegraphics[width=\linewidth]{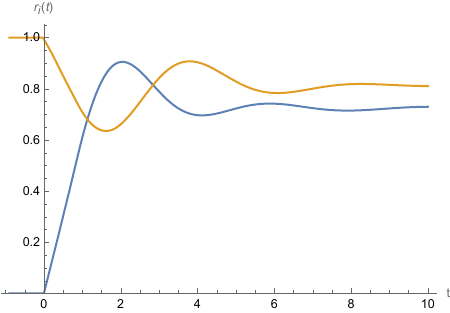}
        \caption{}
        % \label{fig:tau_n_r03}
    \end{subfigure}
    \hfill
    \begin{subfigure}[b]{0.45\linewidth}
        \centering
        \includegraphics[width=\linewidth]{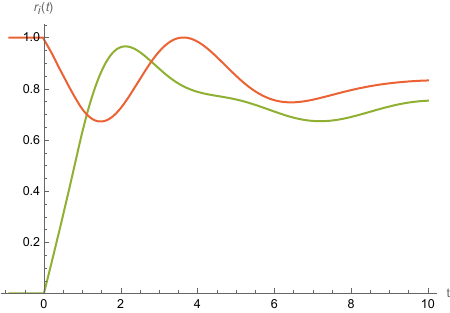}
        \caption{}
        % \label{fig:tau_n_r05}
    \end{subfigure}
    \vfill
    \begin{subfigure}[b]{0.45\linewidth}
        \centering
        \includegraphics[width=\linewidth]{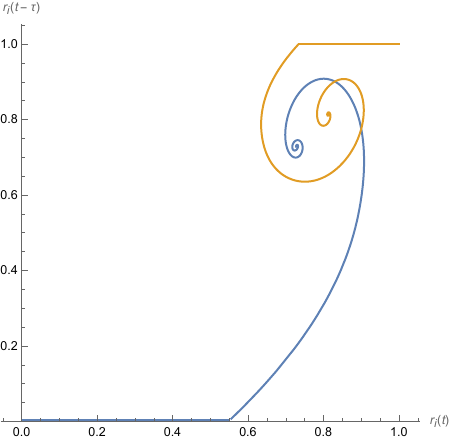}
        \caption{}
        % \label{fig:tau_n_r03}
    \end{subfigure}
    \hfill
    \begin{subfigure}[b]{0.45\linewidth}
        \centering
        \includegraphics[width=\linewidth]{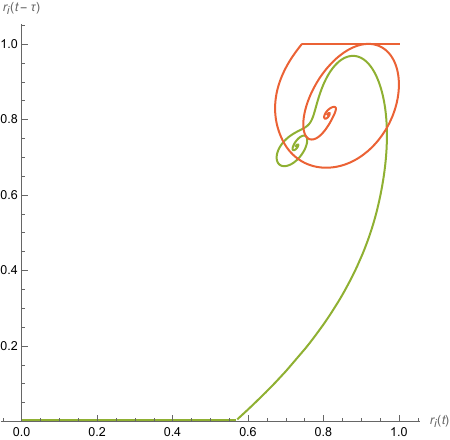}
        \caption{}
        % \label{fig:tau_n_r05}
    \end{subfigure}
    \vfill
    \begin{subfigure}[b]{0.45\linewidth}
        \centering
        \includegraphics[width=\linewidth]{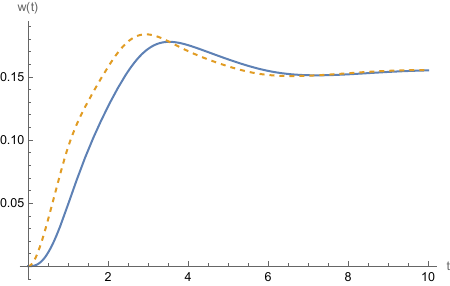}
        \caption{}
        % \label{fig:tau_n_r07}
    \end{subfigure}
    \hfill
    \begin{subfigure}[b]{0.45\linewidth}
        \centering
        \includegraphics[width=\linewidth]{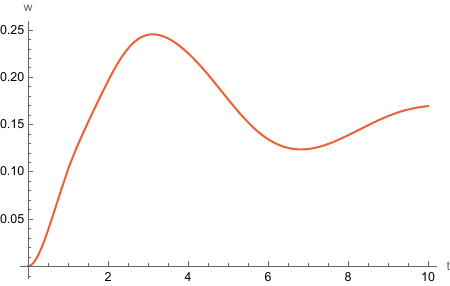}
        \caption{}
        % \label{fig:tau_n_r2}
    \end{subfigure}
    \caption{\centering The plots for the coefficient set $2$ from Table \ref{table:1}.}
    \label{fig:4set}
\end{figure}
Comparing Fig. \ref{fig:4set} to Fig. \ref{fig:3set}, the difference is in the value of the delay $\tau$: it is equal to $\tau=0.9$ and $\tau=0.6$, respectively. The result of this change is that the firing rates have the behavior more resembling periodicity, with larger amplitudes of fluctuations. The time period for which $r_2>r_1$ is much longer compared to Fig. \ref{fig:3set}.
\begin{figure}[H]
    \centering
    \begin{subfigure}[b]{0.43\linewidth}
        \centering
        \includegraphics[width=\linewidth]{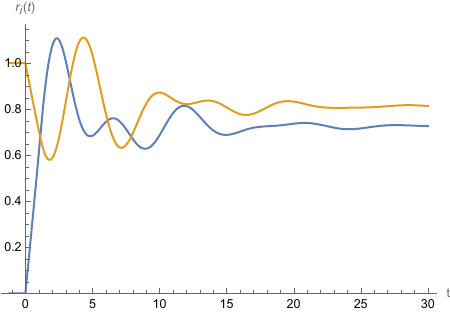}
        \caption{}
        % \label{fig:tau_n_r03}
    \end{subfigure}
    \hfill
    \begin{subfigure}[b]{0.43\linewidth}
        \centering
        \includegraphics[width=\linewidth]{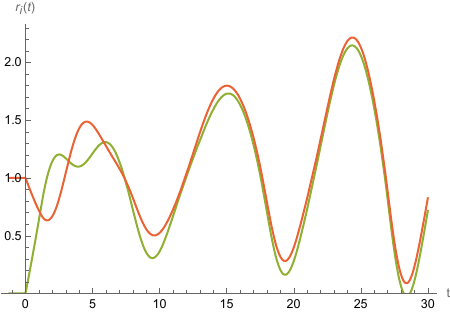}
        \caption{}
        % \label{fig:tau_n_r05}
    \end{subfigure}
    \vfill
    \begin{subfigure}[b]{0.43\linewidth}
        \centering
        \includegraphics[width=\linewidth]{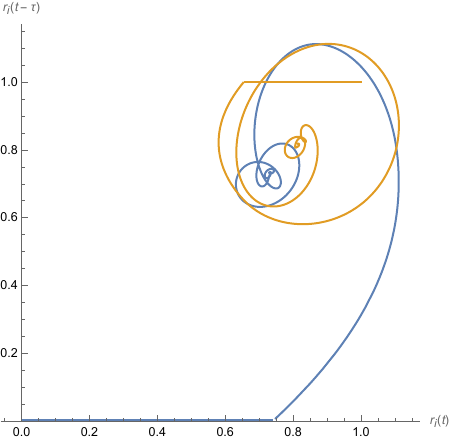}
        \caption{}
        % \label{fig:tau_n_r03}
    \end{subfigure}
    \hfill
    \begin{subfigure}[b]{0.43\linewidth}
        \centering
        \includegraphics[width=\linewidth]{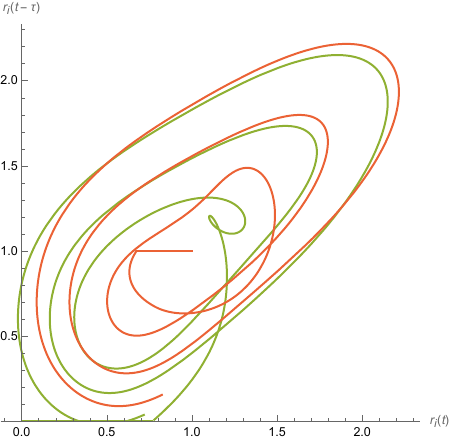}
        \caption{}
        % \label{fig:tau_n_r05}
    \end{subfigure}
    \vfill
    \begin{subfigure}[b]{0.43\linewidth}
        \centering
        \includegraphics[width=\linewidth]{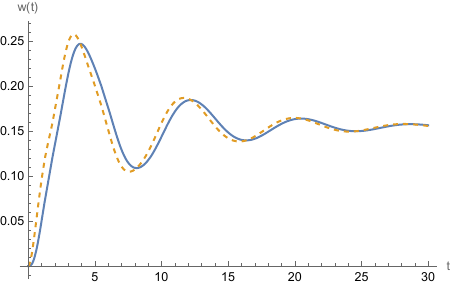}
        \caption{}
        % \label{fig:tau_n_r07}
    \end{subfigure}
    \hfill
    \begin{subfigure}[b]{0.43\linewidth}
        \centering
        \includegraphics[width=\linewidth]{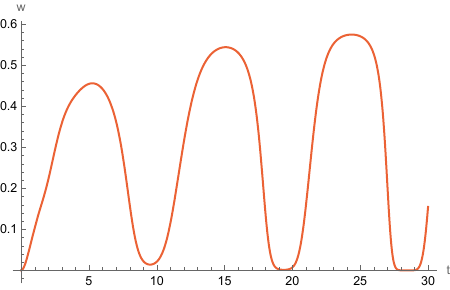}
        \caption{}
        % \label{fig:tau_n_r2}
    \end{subfigure}
    \caption{\centering The plots for the coefficient set $3$ from Table \ref{table:1}.}
    \label{fig:8set}
    \end{figure}
    Compared to Figs. \ref{fig:3set} and \ref{fig:4set}, where $\tau = 0.6$ and $0.9$, respectively, Fig. \ref{fig:8set} differs in the value of the delay, with $\tau = 1.2$. This results in greater amplitudes of solutions' fluctuations. The main novelty is a qualitative difference between the solutions to System \eqref{eq:main3}, which appear to converge to a steady state, and the solutions to System \eqref{eq:mainsystem}, which fluctuate with increasingly larger amplitudes. Around $t=28$ one of the variables becomes negative, therefore losing the biological sense. This shows an interesting consequence of treating the weight $w$ as a separate variable: while the impact of this change might appear insignificant in some of the other figures, it may be the deciding factor when the solutions are close to being unstable.
\begin{figure}[H]
    \centering
    \begin{subfigure}[b]{0.45\linewidth}
        \centering
        \includegraphics[width=\linewidth]{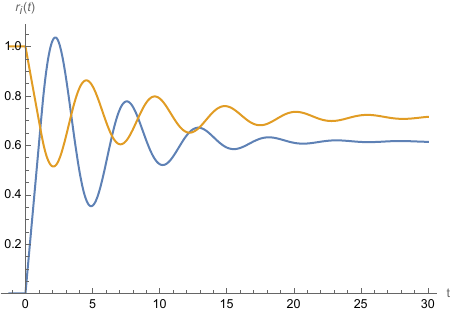}
        \caption{}
        % \label{fig:tau_n_r03}
    \end{subfigure}
    \hfill
    \begin{subfigure}[b]{0.45\linewidth}
        \centering
        \includegraphics[width=\linewidth]{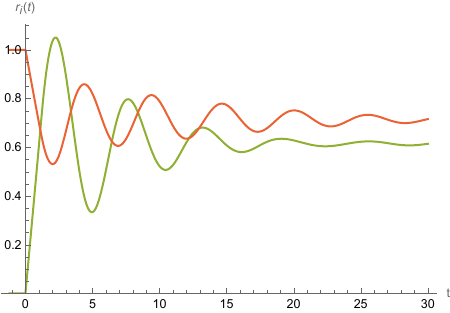}
        \caption{}
        % \label{fig:tau_n_r05}
    \end{subfigure}
    \vfill
    \begin{subfigure}[b]{0.45\linewidth}
        \centering
        \includegraphics[width=\linewidth]{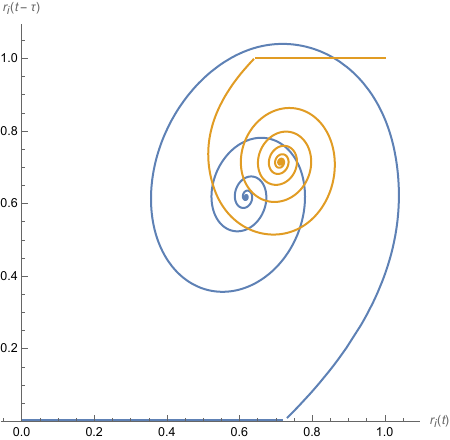}
        \caption{}
        % \label{fig:tau_n_r03}
    \end{subfigure}
    \hfill
    \begin{subfigure}[b]{0.45\linewidth}
        \centering
        \includegraphics[width=\linewidth]{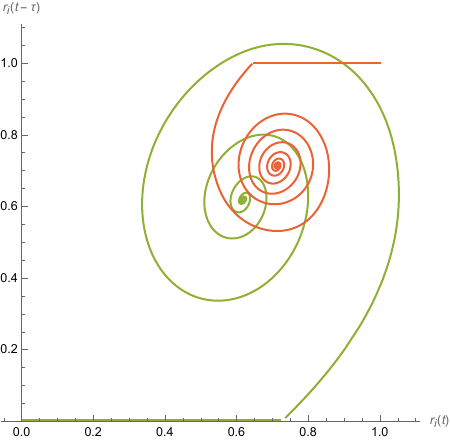}
        \caption{}
        % \label{fig:tau_n_r05}
    \end{subfigure}
    \vfill
    \begin{subfigure}[b]{0.45\linewidth}
        \centering
        \includegraphics[width=\linewidth]{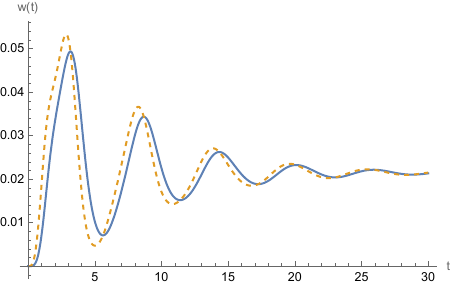}
        \caption{}
        % \label{fig:tau_n_r07}
    \end{subfigure}
    \hfill
    \begin{subfigure}[b]{0.45\linewidth}
        \centering
        \includegraphics[width=\linewidth]{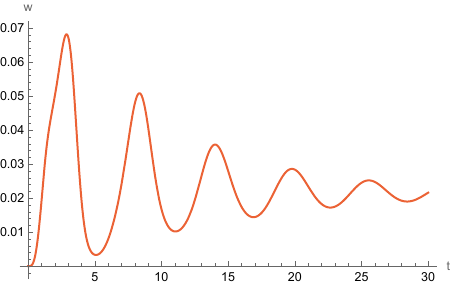}
        \caption{}
        % \label{fig:tau_n_r2}
    \end{subfigure}
    \caption{\centering The plots for the coefficient set $4$ from Table \ref{table:1}.}
    \label{fig:9set}
\end{figure}
The only difference in the coefficients between Figs. \ref{fig:9set} and \ref{fig:8set} is the Hill coefficient, equal to $n=4$ and to $n=2$, respectively. The differences in behaviors of the solutions to the two systems is far smaller in Fig. \ref{fig:9set} -- in particular, they both appear to converge to a~steady state and fluctuate far less.
\begin{figure}[H]
    \centering
    \begin{subfigure}[b]{0.45\linewidth}
        \centering
        \includegraphics[width=\linewidth]{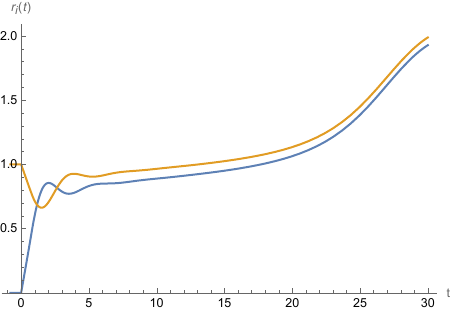}
        \caption{}
        % \label{fig:tau_n_r03}
    \end{subfigure}
    \hfill
    \begin{subfigure}[b]{0.45\linewidth}
        \centering
        \includegraphics[width=\linewidth]{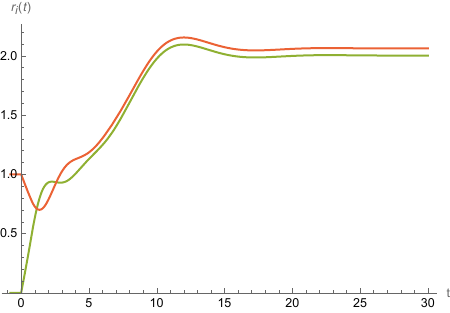}
        \caption{}
        % \label{fig:tau_n_r05}
    \end{subfigure}
    \vfill
    \begin{subfigure}[b]{0.45\linewidth}
        \centering
        \includegraphics[width=\linewidth]{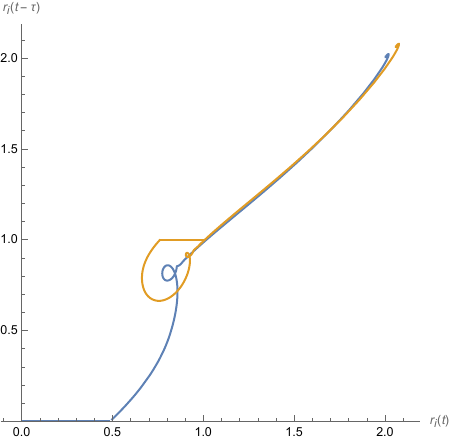}
        \caption{}
        % \label{fig:tau_n_r03}
    \end{subfigure}
    \hfill
    \begin{subfigure}[b]{0.45\linewidth}
        \centering
        \includegraphics[width=\linewidth]{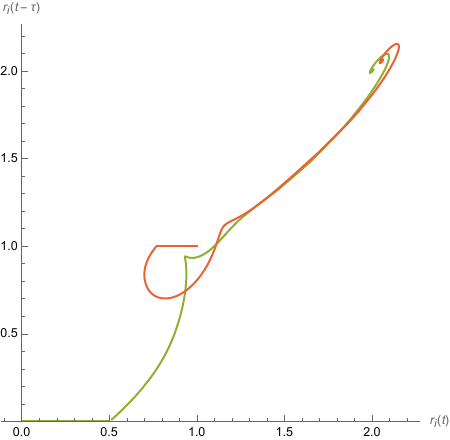}
        \caption{}
        % \label{fig:tau_n_r05}
    \end{subfigure}
    \vfill
    \begin{subfigure}[b]{0.45\linewidth}
        \centering
        \includegraphics[width=\linewidth]{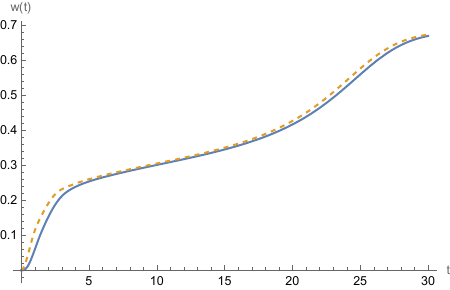}
        \caption{}
        % \label{fig:tau_n_r07}
    \end{subfigure}
    \hfill
    \begin{subfigure}[b]{0.45\linewidth}
        \centering
        \includegraphics[width=\linewidth]{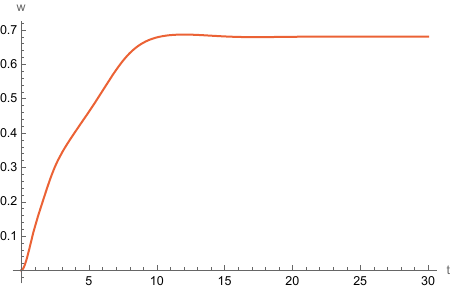}
        \caption{}
        % \label{fig:tau_n_r2}
    \end{subfigure}
    \caption{\centering The plots for the coefficient set $5$ from Table \ref{table:1}.}
    \label{fig:6set}
\end{figure}
The values of the coefficients $\epsilon = 0.72$ and $\tau = 0.8$ in Fig.~\ref{fig:6set} are different than those for previous figures. Compared to Fig. \ref{fig:4set}, the convergence is to the larger stable steady state, while the behavior of the solutions for small $t$ is similar. The plots in Figs.~\ref{fig:6set}(a), (b), (e) and (f) are drawn for $t \leq 30$ and the solutions to the two systems behave quantitatively different, with the ones to System~\eqref{eq:main3} converging to the apparent steady state much later.
\begin{figure}[H]
    \centering
    \begin{subfigure}[b]{0.45\linewidth}
        \centering
        \includegraphics[width=\linewidth]{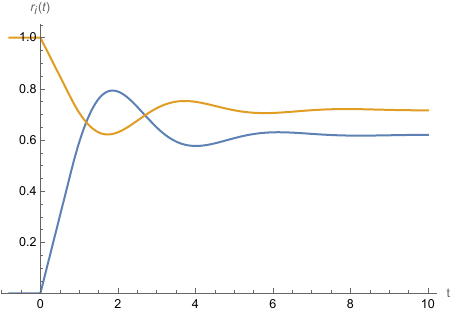}
        \caption{}
        % \label{fig:tau_n_r03}
    \end{subfigure}
    \hfill
    \begin{subfigure}[b]{0.45\linewidth}
        \centering
        \includegraphics[width=\linewidth]{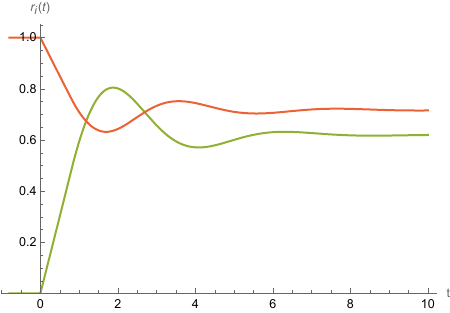}
        \caption{}
        % \label{fig:tau_n_r05}
    \end{subfigure}
    \vfill
    \begin{subfigure}[b]{0.45\linewidth}
        \centering
        \includegraphics[width=\linewidth]{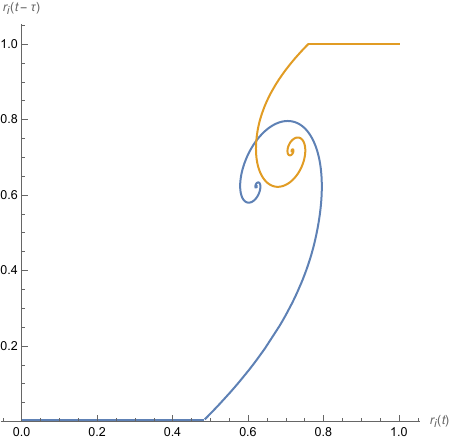}
        \caption{}
        % \label{fig:tau_n_r03}
    \end{subfigure}
    \hfill
    \begin{subfigure}[b]{0.45\linewidth}
        \centering
        \includegraphics[width=\linewidth]{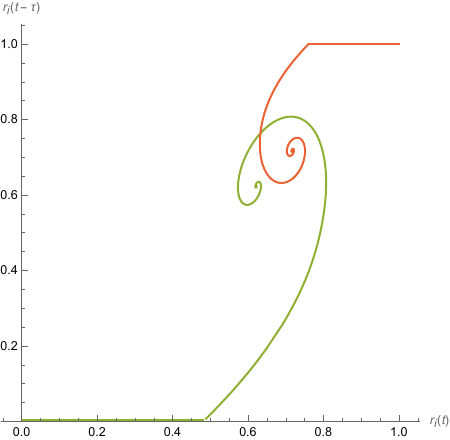}
        \caption{}
        % \label{fig:tau_n_r05}
    \end{subfigure}
        \vfill
    \begin{subfigure}[b]{0.45\linewidth}
        \centering
        \includegraphics[width=\linewidth]{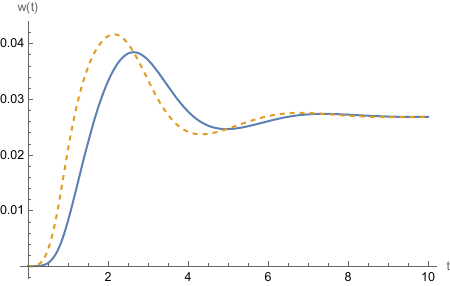}
        \caption{}
        % \label{fig:tau_n_r07}
    \end{subfigure}
    \hfill
    \begin{subfigure}[b]{0.45\linewidth}
        \centering
        \includegraphics[width=\linewidth]{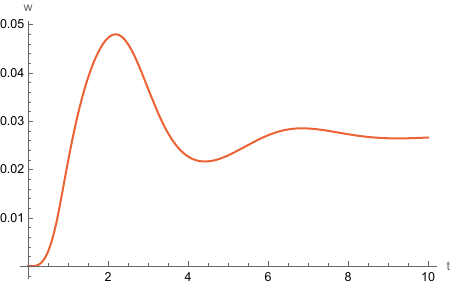}
        \caption{}
        % \label{fig:tau_n_r2}
    \end{subfigure}
    \caption{\centering The plots for the coefficient set $6$ from Table \ref{table:1}.}
    \label{fig:10set}
\end{figure}
The only difference in the coefficients between Figs. \ref{fig:10set} and \ref{fig:6set} is the Hill coefficient, equal to $n=4$ and to $n=2$, respectively.
The main difference in behaviors of the solutions is their convergence to the smaller steady state, instead of a larger one. There is also hardly any distinction between the solutions to the two systems, while in Fig. \ref{fig:6set} it was clear. This result coupled with Fig. \ref{fig:9set} shows that the Hill coefficient not only is able to impact the qualitative behavior of the solutions, but narrows the gap between the two systems' solutions.
\begin{figure}[H]
    \centering
    \begin{subfigure}[b]{0.45\linewidth}
        \centering
        \includegraphics[width=\linewidth]{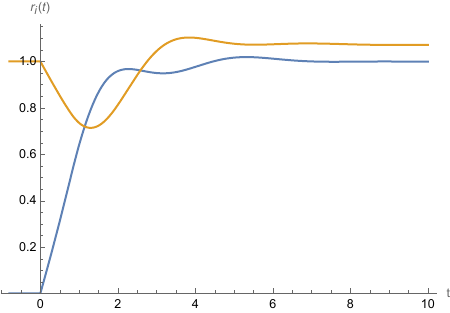}
        \caption{}
        % \label{fig:tau_n_r03}
    \end{subfigure}
    \hfill
    \begin{subfigure}[b]{0.45\linewidth}
        \centering
        \includegraphics[width=\linewidth]{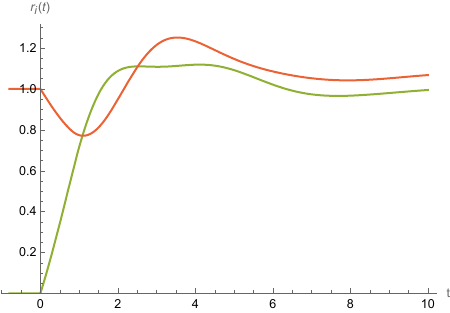}
        \caption{}
        % \label{fig:tau_n_r05}
    \end{subfigure}
    \vfill
    \begin{subfigure}[b]{0.45\linewidth}
        \centering
        \includegraphics[width=\linewidth]{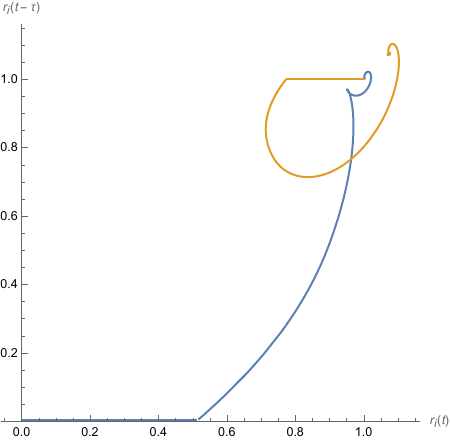}
        \caption{}
        % \label{fig:tau_n_r03}
    \end{subfigure}
    \hfill
    \begin{subfigure}[b]{0.45\linewidth}
        \centering
        \includegraphics[width=\linewidth]{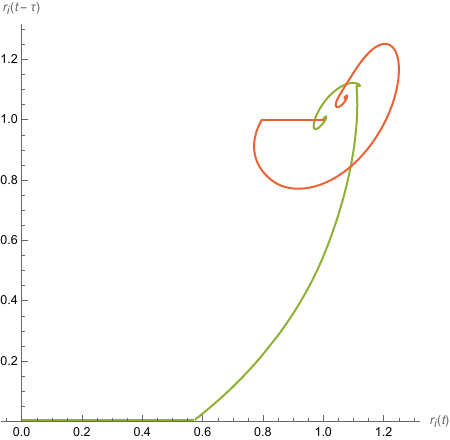}
        \caption{}
        % \label{fig:tau_n_r05}
    \end{subfigure}
        \vfill
    \begin{subfigure}[b]{0.45\linewidth}
        \centering
        \includegraphics[width=\linewidth]{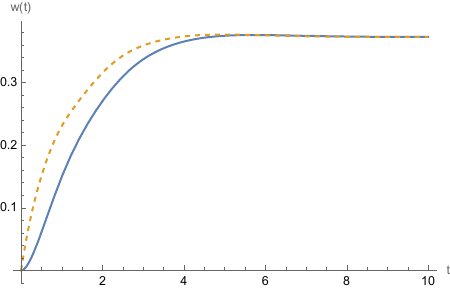}
        \caption{}
        % \label{fig:tau_n_r07}
    \end{subfigure}
    \hfill
    \begin{subfigure}[b]{0.45\linewidth}
        \centering
        \includegraphics[width=\linewidth]{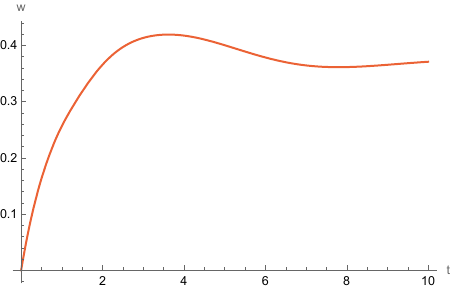}
        \caption{}
        % \label{fig:tau_n_r2}
    \end{subfigure}
    \caption{\centering The plots for the coefficient set $7$ from Table \ref{table:1}.}
    \label{fig:11set}
\end{figure}
The influence of the Hill coefficient $n$ can be seen when comparing Fig. \ref{fig:11set} with Figs. \ref{fig:6set} and \ref{fig:10set}, where $n=1$, $n=2$ and $n=4$, respectively.
%where $n=2$ and $n=4$, respectively, while in Fig. \ref{fig:11set} $n=1$. 
Here, the behavior of the solutions resembles that seen in Fig. \ref{fig:10set}, despite the greater difference in the Hill coefficient $n$. This is likely due to the non-monotonic dependence of the critical delay value $\tau_0$ on $n$.
%This is likely caused by the nonmonotonicity of the critical value of delay $\tau$ with respect to $n$. 
However, note that the gap between the solutions to the two systems is larger and in this sense, the solutions are more akin to those in Fig. \ref{fig:6set}. Also see that the values of the firing rates in the apparent steady state are substantially greater than in Fig. \ref{fig:10set}, but smaller then in Fig. \ref{fig:6set} -- this is likely due to there being only one possible steady state for $n=1$.
\par
%\bdel{Worth noting is the fact that even though the initial condition and the input for firing rate $r_2$ are greater for every coefficient set, for each of them there exists $t>0$ for which $r_1>r_2$} 
As mentioned before, for each of the coefficient sets there exists $t>0$ for which $r_1>r_2$, in spite of greater initial condition and input for firing rate $r_2$.
This cannot be the case for the zero-delay version of the models, which means that the delay introduces ambivalence to the modeled decision-making process.
\par While we do not focus on decision-making itself, this fact shows the importance of the criterion used for this task. For example, the decision made at some arbitrary time $t$ might clearly result in the wrong decision. One could also choose the criterion of the greatest firing rate at any time $t$ -- however, as seen in Figures \ref{fig:9set}(a), (b), this would result in the wrong decision for both models. Another approach is to base the decision on steady state values of the firing rates, as was done in the original model without delays \cite{piskala2017neural}. There are numerous other options to consider, but it is important to keep in mind the significance of this choice.

\section{Discussion}
%\ania{!NOWE! Nie wiem ile ma sensu, to jakaś taka pierwsza wersja, zachęcam do poprawek. W szczególności byłoby super, jakbyś pozmieniał podsumowanie tej całej analizy i dorzucił jakieś wnioski z niej:)}

\noindent In this work, we introduced and analyzed a few variants of the model first proposed by Piskała et al. \cite{piskala2017neural} and then extended by Foryś et al. \cite{forys2017impact} and Bielczyk et al. \cite{bielczyk2019time}. We considered basic properties of the full four-equation model describing the firing rates of two neural populations and the weights of their synaptic connections. Then we analyzed in more detail the behavior of two versions of this model. The first model assumed the synaptic weights to be in their steady state, resulting in two-equation model, while the second one allowed the weights to not be in the steady state but assumed them equal, leaving three equations in the model. In both cases, we started with analyzing the zero-delay model and then introduced delay in self-inhibition of the populations. 

The novelty of this work is the new three-equation model and a successful generalization of considered Hill coefficient values, from $n=2$ used in previous research to a whole set, namely $n \geq 1$.

In the symbolic analysis we showed the existence and stability of the steady states, also indicating the impact of the parameters -- specifically the interplay between the Hill parameter $n$, the maximal synaptic weight $\epsilon$ and symmetric input value $I$ -- on the qualitative behavior of the model. 
%\bdel{Moreover, we investigated the possible loss of stability introduced by the delay, which resulted in Hopf bifurcation in both models with delay.}
Moreover, we investigated the possible loss of stability caused by the increase in the delay and demonstrated that it results in the Hopf bifurcation occurs in both models.
We also compared the behavior of presented models, which turned out to be alike in most cases. We established the existing similarities between them, especially in the zero-delay versions, while also showing the possible differences, for example in the range of attainable critical delay values. Additionally, we used numerical simulations to explore the impact of various parameters, including the delay, on the behavior of the model.

The main limitations of this study include various assumptions which reduce generalizability of the models, e.g. the time-independence of the inputs $I_{1,2}$ or quasi-steady state approximation. We decided to use them for analysis feasibility, however one could consider less constraining assumptions and investigate the model numerically in more detail.
Another limitation is a lack of rigorous theoretical justification of the validity of the quasi-steady state approximation, which could be investigated with the method of perturbation theory -- Tikhonov theorem \cite{Donchev_Slavov_1997}. In general, the solutions of the modified system can be significantly different from the original.

The primary conclusion coming from this study is that various decisions being made in the modeling process, even though can seem inconsequential and be made arbitrarily, in some cases significantly influence the model's behavior. In our case, qualitative behavior of the solutions was dependent on various parameters and the model choice. A comparative and generalizing character of this study provides insight into the impact of such decisions on this decision-making model.

Important point to consider, is that in practice, the values of parameters corresponding to the neural properties should be based on the experimental data, and the rest of the parameters, e.g. the Hill coefficient $n$, should be set to fit the behavior of the real neurons or populations.

\section{References}
\nocite{*}
% Poniżej proszę wpisać pozycje bibliograficzne tak jak w plikach BibTeX (*.bib)

%\selectlanguage{english}
% \bibliographystyle{plain}
%\bibliography{GadekBIB25v2017utf8}
%\bibliographystyle{amsplain}
\bibliographystyle{abbrv}
\bibliography{jobname.bib}
%\centerline{\large\sc References}
%\bibliographystyle{abbrvnat}

% \bigskip
% \setcounter{section}{0}
% \selectlanguage{polish}
% \Polskitrue
% \subjclass{62J05; 92D20}
% \keywords{genetyka statystyczna, wybór modelu, rzadka regresja liniowa, bayesowskie kryterium informacyjne, ilościowa analiza lokalizacji genów}
% * <marek.teuerle@pwr.edu.pl> 2015-07-08T20:15:13.267Z:
%

%\newpage
% \begin{center}
% {\bf Lokalizacja genów.}\\ %The optimal time for a delegation to specialized contractors -- outsourcing --outside-resource-using
%\href{http://wydawnictwa.ptm.org.pl/index.php/matematyka-stosowana/article/view/289/282}{Jan Poleszczuk}
% \href{\repo/597}{Piotr Szulc}
% \end{center}
% \medskip

% \begin{abstract}
% Rozwój genetyki w ostatnich latach doprowadził do sytuacji, w której jesteśmy w stanie przyjrzeć się łańcuchom DNA z dużą precyzją i zebrać ogromne ilości informacji. Oprócz tego okazało się, że zależności między genami a cechami są bardziej skomplikowane niż się wcześniej wydawało. Te dwie rzeczy spowodowały, że niezbędna stała się ścisła współpraca między genetykami a matematykami, których zadaniem jest opracowanie specjalnych metod, radzących sobie w specyficznych i trudnych problemach genetycznych. Artykuł zawiera przegląd zarówno klasycznych jak i najnowszych podejść do problemu lokalizacji genów, czyli wskazywania miejsc w łańcuchu DNA, które istotnie wpływają na interesujące nas cechy. Z powodu nie najlepszej komunikacji między matematykami i genetykami, znajomość metody innych niż klasyczne wśród tej drugiej grupy jest wciąż niewielka.
% \end{abstract}

\Koniec
\end{document}